%% file: paper.tex
\PassOptionsToPackage{table, x11names, dvipsnames, svgnames}{xcolor}
\documentclass[final,3p,twocolumn]{elsarticle}
\pdfoutput=1
\usepackage[UKenglish]{babel}
\usepackage[T1]{fontenc}
\usepackage{url}
\usepackage{graphicx}
\usepackage{svg}
\usepackage[export]{adjustbox}
\usepackage{amsmath}
\usepackage{amsfonts}
\usepackage{xcolor}
\usepackage{numprint}
\usepackage{csquotes}
\usepackage{tabularx}
\usepackage{booktabs}
\usepackage{siunitx}
\usepackage{tikz}

\usepackage[breaklinks]{hyperref}

\usepackage{multirow}
\usepackage{soul}

\usepackage{breakurl}

\usepackage{placeins}
\usepackage{verbatim}

\usepackage[inline]{enumitem}
\setlist*[enumerate]{label=(\arabic*)} 

\DeclareMathOperator*{\argmax}{arg\,max}

\newcommand{\website}{\href{https://cdn.lfdr.de/vq-dqn}{website}}

\usetikzlibrary{quantikz}

\catcode`&=4

\ExplSyntaxOn\makeatletter
\RenewDocumentEnvironment{quantikz}{O{} +b}
 {
  \tl_set:Nn \l__quantikz_body_tl { \tikzcd@ [#1] #2 }
  \regex_replace_all:nnN { \& } { \cA\& } \l__quantikz_body_tl
  \tl_use:N \l__quantikz_body_tl
  \endtikzcd
 }{}
\tl_new:N \l__quantikz_body_tl
\makeatother\ExplSyntaxOff

\begin{document}

\begin{frontmatter}
\title{Uncovering Instabilities in Variational-Quantum Deep Q-Networks}
\author[1]{Maja Franz\fnref{equiv}}
\ead{maja.franz@othr.de}
\author[1]{Lucas Wolf\fnref{equiv}}
\ead{lucas.wolf@st.othr.de}
\author[2]{Maniraman Periyasamy\fnref{equiv}}
\ead{maniraman.periyasamy@iis.fraunhofer.de}
\author[2]{Christian Ufrecht}
\ead{christian.ufrecht@iis.fraunhofer.de}
\author[2]{Daniel D.\ Scherer}
\ead{daniel.scherer2@iis.fraunhofer.de}
\author[2]{Axel Plinge}
\ead{axel.plinge@iis.fraunhofer.de}
\author[2]{Christopher Mutschler}
\ead{christopher.mutschler@iis.fraunhofer.de}
\author[1,3]{Wolfgang Mauerer}
\ead{wolfgang.mauerer@oth-regensburg.de}

\fntext[equiv]{These authors contributed equally (name order randomised).}
\address[1]{Technical University of Applied Sciences Regensburg, Germany}
\address[2]{Fraunhofer IIS, Fraunhofer Institute for Integrated Circuits IIS, Division Positioning and Networks, Nuremberg, Germany}
\address[3]{Siemens AG, Corporate Research, Munich, Germany}

\begin{abstract}

Deep Reinforcement Learning (RL) has considerably advanced over the past decade. At the same time,
state-of-the-art RL algorithms require a large computational budget in terms of training time to converge.
Recent work has started to approach this problem through the lens of quantum computing, which promises
theoretical speed-ups for several traditionally hard tasks. In this work, we examine a class of hybrid
quantum-classical RL algorithms that we collectively refer to as variational quantum deep Q-networks (VQ-DQN).
We show that VQ-DQN approaches are subject to instabilities that cause the learned policy to diverge, study
the extent to which this afflicts reproduciblity of established results based on classical simulation, and perform
systematic experiments to identify potential explanations for the observed instabilities. Additionally, and in contrast to most
existing work on quantum reinforcement learning, we execute RL algorithms on an actual quantum processing unit
(an IBM Quantum Device) and investigate differences in behaviour between simulated and physical quantum systems
that suffer from implementation deficiencies. Our experiments show that, contrary
to opposite claims in the literature, it cannot be conclusively decided if known quantum approaches,
even if simulated without physical imperfections, can provide an advantage as compared to classical
approaches. Finally, we provide a robust, universal and well-tested implementation of
VQ-DQN as a reproducible testbed for future experiments.

\end{abstract}

\end{frontmatter}

\section{Introduction}

Techniques for reinforcement learning (RL) have seen considerable progress during the past decade. Driven by both, algorithmic advances and the emergence of deep learning \cite{goodfellow16, murphy12, lecun15}, RL has emerged from a conceptual approach  to successfully tackling tasks previously deemed infeasible. This includes aspects of robotic manipulation~\cite{Levine16, Hoof16, openai18, Kalashnikov18}, autonomous driving~\cite{Bhalla20, Baheri20, Huang21}, and mastering combinatorially-hard board
games~\cite{silver16, silver17a, silver17b, Schrittwieser19}. At the same time, state-of-the-art deep RL methods require an exorbitant computational budget to match or exceed human performance on seemingly simple tasks, such as playing arcade video games. As an example, Badia et al.~\cite{Badia20} report training times of roughly \numprint{53000} hours, distributed over 256 machines, to achieve superhuman performance on all 57 Atari games of the Arcade Learning Environment benchmark \cite{ale}. Also, the learning dynamics of these approaches, both in terms of stability and optimality, are not yet fully understood and remain a subject of current research~\cite{doubledqn, Hasselt18, Ilyas18, Agarwal19}.

Concurrent to these developments, quantum computing~\cite{nielsen16} has started to receive increasing interest in real-life applications. It promises computational
speedups, especially selected weakly-structured search problems like integer factoring \cite{Shor99}, or exploration of unstructured search spaces \cite{Grover96, Grover98} by exploiting fundamental phenomena of quantum mechanics (see Sec.~\ref{subsec:vqc}). 
Reinforcement learning can be regarded as a search problem (in terms of seeking an optimal policy, as we outline in Sec.~\ref{subsec:dqn}). Consequently, it is natural to ask whether a quantum speedup is realisable in this domain. 

Limitations on achievable speedups have been studied in detail~\cite{StilckFranca2021}, and lower bounds are known for several
important fundamental problems~\cite{Buhrman:2021}. Despite numerous technological challenges rooted in, amongst others, noise and imperfections of near-term intermediate scale quantum devices~\cite{Preskill2018quantumcomputingin}, sufficient margins
for industrially relevant improvements remain~\cite{bayerstadler2021,bova2021},
but necessitate a more precise understanding and a critical evaluation of the performance
of quantum approaches on currently available hardware designs. Since 
RL, like other machine learning approaches, relies on stochastic components 
that may amplify variations in algorithmic performance
(and, more generally, challenge replication efforts), this is another aspect that
requires careful consideration.

In this article, we examine and extend a class of recent hybrid quantum-classical approaches to
reinforcement learning that we collectively refer to as \emph{Variational-Quantum Deep Q-Networks}
(VQ-DQN). Originally proposed by Chen et al.~\cite{vqdqn} and later refined by Lockwood and
Si~\cite{Lockwood2020}, VQ-DQN builds upon the deep Q-networks (DQN) algorithm \cite{dqn13, dqn15}, which
replaces the core neural network component with a quantum machine learning model, namely, a variational
quantum circuit (VQC)~\cite{Mitarai18}. Although the results published in \cite{dqn13, dqn15} promise
interesting properties, we show that VQ-DQN approaches are subject to instabilities that ultimately cause
the learned policy to diverge. Policy divergence is obviously detrimental to the practical utility
of the approach, especially if it already happens in perfect simulations of quantum systems.
Quantum computers that can be manufactured under the constraints of current technological limitations
additionally suffer from noise, imperfections, and very strongly limited amounts of 
available quantum bits. They are referred to as \emph{noisy, intermediate scale quantum
computers} (NISQ). To understand the additional degradation caused by these imperfections on the
performance of RL approaches, we perform comparative experiments on actual quantum
hardware---a gate-based IBMQ device 
(\href{https://www.ibm.com/blogs/research/2020/07/qv32-performance/}{Falcon r4}) 
operated in Ehningen, Germany.

In general, our investigation is part analysis and part reproduction study,
and we provide a reproduction package with a well-tested implementation\footnote{See
\url{https://doi.org/10.5281/zenodo.7030069}} of
VQ-DQN. To make best use of
available libraries and to provide an open testbed for future experiments, our implementation
is written in two separate quantum frameworks, which are each coupled to a machine-learning
framework: Tensorflow-Quantum~\cite{Broughton2021}(TFQ)/Tensorflow~\cite{Abadi15} and
Qiskit~\cite{Abraham19}/Pytorch~\cite{Paszke2019}.

The paper is structured as follows: Section \ref{sec:background} provides a concise introduction to DQN (\ref{subsec:dqn}), VQCs (\ref{subsec:vqc}), and the VQ-DQN algorithm (\ref{subsec:vqdqn}). Section \ref{sec:related-work} reviews related work. Section \ref{sec:reproduction} describes our methodological approach towards finding and characterising instabilities. Section \ref{sec:experiments} summarizes our experiments. Section \ref{sec:ibmq_validation} explains the validation experiment on real quantum device. Further, we proceed to compare the DQN with variational quantum circuit against a DQN with classical neural network in Section \ref{sec:comparison}.  Finally, we conclude in Section \ref{sec:conclusion}.

\section{Background}
\label{sec:background}
To introduce the concepts used in this study, the following paragraph discusses notation
and basic principles of both, machine learning and quantum computation. 

\subsection{Deep Q-Learning}
\label{subsec:dqn}

Most formulations of RL center around the notion of a \textit{Markov decision process} (MDP)~\cite{Sutton99}, where an \emph{agent} interacts with an \emph{environment} at discrete time steps $t$. In each time step, the current configuration of the environment is summarised by the \textit{state} $S_t \in \mathcal{S}$. Based on this information, the agent selects an \textit{action} $A_t \in \mathcal{A}$ according to a \textit{policy} $\pi(s,a) = \mathbb{P}[A_t = a| S_t = s]$. Executing the selected action causes a transition of the environment to a next state $S_{t+1}$; simultaneously, the agent receives a scalar \textit{reward} $R_{t+1}$ that quantifies the contribution of the 
selected action towards solving the task. The agent's goal is to maximize the return, i.e., the discounted sum of rewards, $G_t = \sum_{t'=t}^T \gamma^{t'} R_{t'}$  until a terminal state $S_T$ is reached. In that, the discount factor $\gamma$ controls how much the agent favors immediate over future rewards. Both $S_{t+1}$ and $R_{t+1}$ are assumed to obey the Markov property (i.e., conditional independence of previous states and actions given $S_t, A_t$). However, the MDP's \textit{dynamics}, $\mathbb{P}[S_{t+1}, R_{t+1} |  S_t, A_t]$, are typically unknown to the agent, which necessitates learning a policy by trial-and-error.

The fundamental idea of \textit{Deep Q-Learning} (also referred to as \textit{deep Q networks}, DQN) \cite{dqn13, dqn15} is to learn the \textit{optimal state-action value function} $Q_*(s,a) = \max_\pi \mathbb{E} \left[G_t | S_t = s, A_t = a, \pi \right]$ 
-- that is, the return expected when taking action $a$ in state $s$, and then following an optimal policy in all future states. Once $Q_*(s,a)$ is known, an optimal policy can be easily recovered by selecting actions greedily, that is  $\pi_*(s) = \argmax_{a} Q_*(s,a)$. This is achieved by training a neural network to satisfy the well-known \textit{Bellman Optimality Equation} (BOE) that relates the values of a state-action pair to the value of the next state:
\begin{align}
    \hspace*{-1em}Q_*(s,a) = \mathbb{E}\Big[R_t + \gamma \max_{a'} Q_*(S_{t+1}, a') \mid &S_t = s, \nonumber\\
                                                                         &A_t = a\Big]\label{eq:boe}
\end{align}

More concretely, the deep Q-network is trained to minimize the difference between the left- and right-hand side of this equation (also known as the \textit{temporal difference error} or \textit{TD-error}), under some loss function (e.g., $L_2$), evaluated on mini-batches of transitions $(S_t, A_t, R_{t+1}, S_{t+1})$ sampled by the agent. These transitions are sampled using an \textit{off-policy} approach -- instead of applying the current greedy policy (also called \textit{target policy}), an $\epsilon$-greedy \textit{behavior policy} that selects a random action with probability $\epsilon$ is chosen. Decaying $\epsilon$ over the course of training allows the agent to explore the environment, while guaranteeing that the behavior policy and target policy (and hence, the underlying data distributions) converge eventually.

As Mnih et al.~\cite{Mnih16} point out, learning $Q_*$ with a high-capacity function approximator leads to convergence problems. To this end, DQN makes use of (1) a \textit{target network}, which is a copy of the deep Q-network with temporarily fixed weights to evaluate the right-hand side of \ref{eq:boe}, and (2) an \textit{experience replay buffer} \cite{Lin92} from which experienced transitions are re-sampled for mini-batch gradient descent. For a detailed discussion of these specifics, we refer the interested reader to \cite{dqn13, dqn15}.

\subsection{Variational Quantum Circuits}
\label{subsec:vqc}

Quantum computation uses the \textit{qubit} as the fundamental unit of information. In contrast to classical bits, a set of $n$ qubits can not only assume the $2^n$ classical basis states (i.e., $0, 1, \dots, 2^n -1$), but also any superposition of these basis
states. Note that superimposable quantum states reside in an infinite state space than their classical counterparts, which is often seen as an
indication of increased computational capabilities, although the exact reason
for possible quantum speeds remains elusive~\cite{arute2019quantum}.

The variational quantum circuit is a machine learning model based on quantum circuits~\cite{Mitarai18}. Similar to neural networks, VQCs consist of sequential \textit{layers} that represent parameterised transformations on the VQC's quantum state. In particular, VQC layers apply e.g. learnt single-qubit rotations (in \(X\)-, \(Y\)-, and \(Z\) direction using
the corresponding Pauli operators~\cite{nielsen16}) to each qubit of the circuit. Entanglement can be generated by applying a series of CNOT-gates~\cite{nielsen16} to pairs of qubits. The specific single-qubit rotation parameters are learned via gradient-descent on an error signal, computed over the expected measurements in \(Z\) direction of one or more output qubits.

\subsection{VQ-DQN}
\label{subsec:vqdqn}


Variational quantum deep Q-networks (VQ-DQN)~\cite{vqdqn, Lockwood2020, skolik2021}
replace the deep neural network in DQN with a VQC.

\subsubsection{Q-value extraction}
 \label{sec:extraction}
For a given input MDP state, Q-values are predicted for all $|\mathcal{A}|$ actions 
simultaneously by taking the expectation value of a measurement (in \(Z\) direction) of a 
corresponding number of output qubits. The resulting measurements lie within $[-1;1]$;
obtaining valid action values thus requires further processing, for instance by scaling the
measured results by a learnt multiplicative factor. 

\subsubsection{Input encoding}\label{sec:input-enc}
To input a (classical) MDP state $s \in \mathcal{S}$ to the VQC, that state needs to be represented as a
quantum state $\vert \Psi(s) \rangle$ using the available qubits. Chen et al.\ \cite{vqdqn} address this problem by 
only considering MDPs with discrete state spaces and associating each MDP state with one of the $2^N$
quantum basis states. Lockwood and Si \cite{Lockwood2020} and Skolik et al.\  \cite{skolik2021} extend this method to MDP states with continuous 
components with a simple encoding scheme, with which the authors report results on the ``Blackjack'' 
and ``CartPole-v0'' environments (see Ref.~\cite{Brockman16} for implementation details).
In particular, each component of the input state $s$ is encoded by applying parameterised Pauli rotation 
gates~\cite{nielsen16} to one respective qubit in the circuit (initialised to $\vert 0 \rangle$). Lockwood and Si~\cite{Lockwood2020} 
propose two encoding schemes: \textbf{Scaled (S)} encoding, which determines a rotation angle by
scaling finite-domain input components to $[0, 2\pi]$, and \textbf{Directional (D)} encoding, which
encodes infinite-domain inputs by rotating the qubit by $\pi$ if the input is greater than $0$. 
Skolik et a.~\cite{skolik2021} additionally present \textbf{Continuous (C)} encoding, which
computes rotation angles as the $\arctan$ of the respective input component.

\section{Related Work}
\label{sec:related-work}

\subsection{Deep Q-Learning and its instabilities}

The DQN approach dates back to Watkin's Q-Learning \cite{WatkinsDayan92} and has seen a lot of interest over the years due to its immense potential in learning capabilities. Deep Q-Learning is itself an active field of research because of its versatility in end applications.  Nevertheless, as versatile as the end applications are, the algorithm possesses space for improvements in its stability and speed of convergence to a solution \cite{doubledqn, duelingdqn, prioritizedreplay, Horgan18, gorilladqn, ngu, r2d2, Badia20}.  In particular, the Q-learning approaches, i.e., off-policy learning with function approximation and bootstrapping,  are known to diverge in certain scenarios. This divergence occurs more often when the Q-value is approximated using a non-linear function approximator such as a deep neural network. However, the root causes are still unknown \cite{Tsitsiklis97, Hasselt10, Sutton15, Hasselt18}.

\subsection{Quantum Reinforcement Learning}

Over the past few years, there have been several attempts to improve the performance of reinforcement learning algorithms via possible `quantum advantage' using quantum computing. Like in the classical realm, no one method has emerged as the superior approach in performance or generality. The first quantum reinforcement learning (QRL) algorithm (to our knowledge) has been proposed by Dong et al.\ \cite{Dong08}, which uses a modified version of Grover's algorithm \cite{Grover96} to learn a state-value function. As in the classical reinforcement learning family, whose members vary in algorithm and methodology, various algorithms for QRL have been studied ~\cite{Dunjko15, Flamini19, Neukart17, Silver14}.  The VQ-DQN algorithm was originally proposed by Chen et al.\ \cite{vqdqn} where the authors have used variational quantum circuits to solve two different discrete environments, namely, `cognitive radio' and `frozen lake'. Both these environments are discrete environments where the state space is finite. The next study on VQ-DQN algorithm was conducted by Lockwood and Si\ \cite{Lockwood2020}, where the authors used a VQC to solve both continuous and discrete environments.  Another study that analyses the learning performance and behavior of VQ-DQN was conducted by Skolik et al.\ \cite{skolik2021}. Here the authors explore the effects of having a VQC as a Q-value approximator along with techniques like data re-uploading and a hybrid quantum-classical model.

\section{Reproduction study}
\label{sec:reproduction}

\begin{figure}[ht]
    \centering
    \vspace*{0em}\input{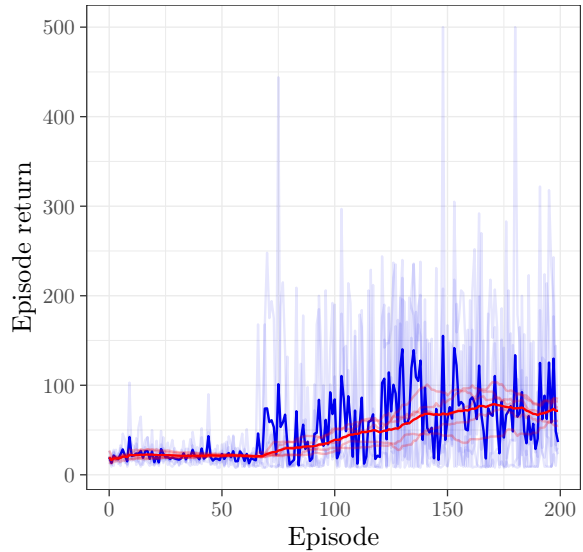}\vspace*{-1.7em}
    \caption{Results from Ref.~\cite{Lockwood2020}, reproduced using the published source code. The light blue lines indicate the total reward collected in an episode, using the greedy policy for each agent. The light red lines represent a moving average of the (up to) 20 previous episode returns. Results are averaged over five experiments, which is represented by the strong red and strong blue lines. The experiments are based on the \texttt{CartPole-v1} environment, where the maximum achievable return value is \numprint{500}.}
    \label{fig:lockwood-cartpole-reproduction}
\end{figure}

To gauge the learning capability of VQ-DQN, we first reproduce the results published by Lockwood and
Si~\cite{Lockwood2020} on the \texttt{CartPole-v1} task (cf.~Sec.~\ref{subsec:vqdqn}). We train five
VQ-DQN agents and evaluate their performance during training using the source code\footnote{Available on
\href{https://github.com/lockwo/quantum_computation/commit/0c702f92bb7e8519073b04a5247ba9e0a34e0198}{GitHub}
(link in PDF).} published by the authors. The results are visualised in
Fig.~\ref{fig:lockwood-cartpole-reproduction}. The blue line indicates episode returns. The red line represents
a moving average of the (up to) 20 previous returns.\footnote{Note that these statistics have been measured with
the original source code, without modification. Superficial differences in visual appearance are caused by the
plot aesthetic settings.} While our measurements reproduce the computational outcome of the published results,
we identify two notable methodological aspects that require careful consideration and interpretation: 

    \textbf{Training frequency}---A step of mini-batch gradient descent is carried out only once per episode (namely, after its termination). This differs substantially not only from the original
    DQN algorithm, but also from the pseudo-code provided by Lockwood and Si \cite{Lockwood2020}, were training is executed in regular intervals after a set number of trajectories has been sampled by the agent. We are not aware of other approaches in the literature that pursue or analyse this approach, and conjecture that it might have a detrimental effect on learning, since the distribution of transitions in the replay buffer grows faster than the amount of data that the agent perceives. The adaptation also complicates the comparison between independent runs of the algorithm, depending on the length of the experienced episodes.

    \textbf{Performance evaluation}---Measuring agent performance in terms of a moving average over previous runs is not a good indicator for learning success: Averaged returns have been generated by different policies, that is, trained on increasing numbers of transitions at different stages of $\epsilon$-decay. Further, the averaging approach shadows any underlying instabilities as indicated by the raw episode returns: In all five runs, the blue line oscillates strongly between low and high return values, indicating that the underlying policy network/circuit fails to converge towards an optimal policy. Note that in complex environments, DQN convergence \textit{can} be non-monotonic in terms of measured returns (see, e.g., Ref.~\cite{dqn13}). Observing oscillations of this magnitude on \texttt{CartPole} (which can be learnt in an approximately monotonic fashion by a simple neural network with DQN, refer to ~\ref{sec:comparison}) does not give a promising outlook on VQ-DQN's capability to generalise to more challenging tasks. \newline

Besides, we would like to explicitly point out that the experiment is based on \texttt{CartPole-v1},
where return values of up to 500 can be achieved. In contrast, returns in 
\texttt{CartPole-v0} cannot exceed 200, which is important to take into account when
judging closeness to optimality of particular approaches, especially when the visual
display of episode return time series uses clipped axes.

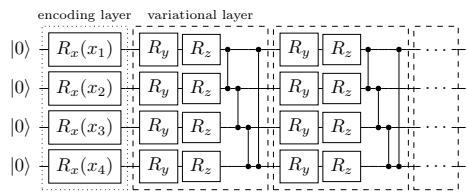
\begin{figure}[tbph!]
\centering
\input{img/vqc_standard}
\caption{Standard VQC architecture}
\label{fig:vqc_standard}
\end{figure}

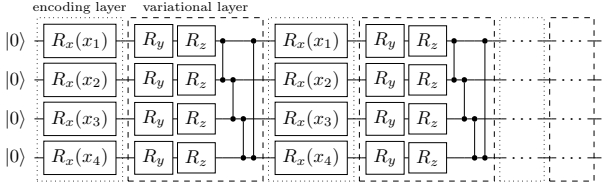
\begin{figure}[tbph!]
\centering
\input{img/vqc_data_reuploading}
\caption{VQC architecture with data re-uploading strategy.}
\label{fig:vqc_data_reuploading}
\end{figure}

One other study which overcame these instabilities using a VQ-DQN algorithm to solve the Cartpole environment is conducted by Skolik et al.  \cite{skolik2021}. Here the authors have used slightly different gate connectivity in their VQC compared to Lockwood and Si~\cite{Lockwood2020}. Apart from the change in VQC architecture, the authors also perform a gradient descent optimization step after every 30 sampling steps. They also present their total reward attained in each episode averaged over ten different agents rather than presenting a moving average.
\newline
Skolik et al. \cite{skolik2021} have studied and tested various combinations of pure and quantum-classical hybrid VQC architectures in their work. However, the pure VQC model did exhibit the same instabilities exhibited by Lockwood and Si's model. Skolik et al. \cite{skolik2021} used a hybrid VQC model where the inputs to and outputs from the VQC were multiplied with classical weights' along with the data re-uploading strategy~\cite{salinas2020} to overcome these instabilities. Data re-uploading is a strategy where the encoding circuit is reintroduced at multiple instances in a VQC.
The standard encoding method follows a traditional neural network setup where the input to the network generally comes before the variational layers as shown in figure \ref{fig:vqc_standard}. However, in a gate-based VQC, both the input and the variational parameters are fed into the circuit as rotational angles. Therefore, there is no theoretical limitation on the maximum number of gates nor the number of repetitions of input features that can be fed into the circuit. Hence, the encoding circuit can be placed before every variational layer as shown in figure \ref{fig:vqc_data_reuploading}. 

Reintroducing the encoding circuit increases the expressivity of the model~\cite{Schuld2021}. It was shown by Schuld et al.~\cite{Schuld2021} that the functions represented by VQCs are Fourier sums.  In which, the variational layers determine the amplitudes of the Fourier sum and the encoding layer fixes the frequency spectrum. Hence, the more encoding layers present via data re-uploading, the larger the frequency spectrum represented by the VQC the higher the expressivity of the represented function class can be. Even though the hybrid model exhibited a relatively stable learning behavior, the impact of classical weights on the overall training process is not distinguished nor studied. 
The results of our replication attempt of the work by Skolik et al.~\footnote{The associated source code published by the authors of \cite{skolik2021} is available on \href{https://github.com/askolik/quantum_agents}{GitHub}(link in PDF). Skolik et al. also provide a simplified implementation as a tutorial in the \href{https://www.tensorflow.org/quantum/tutorials/quantum_reinforcement_learning\#3_deep_q-learning_with_pqc_q-function_approximators}{TFQ documentation}(link in PDF). Note that we were not aware of these implementations during our reproduction process.}
are shown in Fig.~\ref{fig:skolik_reproduction}. These experiments were conducted based on the parameters given in the Appendix section of Ref. \cite{skolik2021}. The measurement results shown in Fig.~\ref{fig:skolik_reproduction} confirm the published results. 

\begin{figure}[ht]
    \centering
    \vspace*{0em}\input{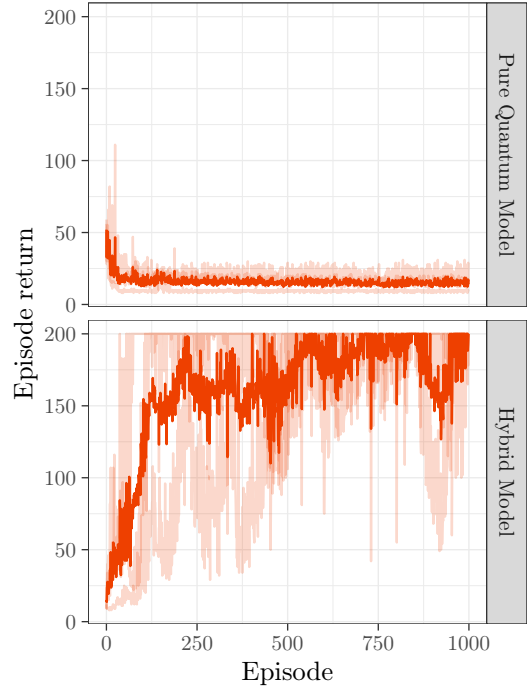}\vspace*{-1.7em}
    \vspace{-1em}
    \caption{Replication attempt of the results from \cite{skolik2021}. Here we replicated the hybrid quantum-classical model with data re-uploading and the pure quantum model with data re-uploading as proposed by Skolik et al.~\cite{skolik2021}. The experiments are based on the \texttt{CartPole-v0} environment, where the maximum achievable return value is \numprint{200}.}
    \label{fig:skolik_reproduction}
\end{figure}

\section{Experiments}
\label{sec:experiments}
Previous implementations of the VQ-DQN approach show various methodological
issues~\cite{Lockwood2020} that we have discussed in detail in the previous section.
For having a stable and uniform VQ-DQN framework that coincides with the classical RL
practices and to provide a replication of existing results on top of mere 
reproduction, 
we re-implement the original deep Q-learning algorithm as described in \cite{dqn13, dqn15} 
in Tensorflow~\cite{Abadi15}/
Tensorflow-Quantum~\cite{Broughton2021}(TFQ). In contrast to the previous implementations, which use TFQ too, our re-implementation allows to conventiently integrate extensions and has a higher degree of configurability of hyperparameters. Furthermore we included a flexible validation mechanism, which is used to evaluate the performance of a current policy. Since in previous implementations a fair comparison between different studies was difficult due to several meanings of return values (e.g. averaging over past episodes as in Ref.~\cite{Lockwood2020} vs. taking a single episodes return value as in Ref.~\cite{skolik2021}), we designed our validation mechanism to allow a uniform comparison of different classical and quantum RL approaches.
(Sec.~\ref{sec:methodology} discusses implementation details).
This section covers experiments, which were conducted using the quantum simulators of the TFQ framework. In addition to our TFQ implementation, we also ported the code to the Qiskit~\cite{Paszke2019} framework in order to run experiments on the IBM Quantum~\cite{ibmq2021} devices, which is described in detail in sec.~\ref{sec:ibmq_validation}

Using our TFQ-implementation, we run a set of experiments to systematically evaluate the observed
instabilities. Throughout all our experiments, we used the \texttt{CartPole-v0} environment to ensure comparability with ~\cite{skolik2021} and \cite{Lockwood2020}, and also to keep computational cost at bay. Sec.\ \ref{sec:encoding-extraction} investigates the effects of the chosen input encoding and Q-value extraction method on performance and stability. Using these insights, we run an extensive cross-validation study described in Sec.\
\ref{sec:cross-validation}. Additionally we have investigated properties of the VQC parameter space as a potential cause for instabilities; 
as the experiments conducted based on this speculation did not lead to a justifiable root cause, we focus only on the experiments on the input-encoding, Q-value extraction methods, and cross-validation mentioned above in this paper. However, we have included a brief discussion in \ref{sec:appendix_a} for reference.

\subsection{Methodology}
\label{sec:methodology}
To describe our methodology, let us first set the employed conventions: By \emph{sampling steps},
we refer to the transitions sampled from the $\epsilon$-greedy behavior policy. By \emph{training step}, we
understand one iteration of gradient descent. Words in \texttt{monospaced} font indicate configurable
parameters of the algorithms.

To ensure comparability between our different experimental setups, and
especially between previous research and our dedicated experiments, we choose
sampling steps as fundamental unit of training time.  Each experiment is run for \numprint{50000} sampling
steps. We deliberately use a long time horizon to capture any phenomena that may materialise late in the
learning process caused by slow convergence, but retain the possibility to terminate successful runs
prematurely, as described in detail below. Initially, the replay memory is pre-filled with
\texttt{train\_after=1000} sampling steps, corresponding to at least five full episodes, using a uniform random policy with $\epsilon=1$.  

A sampling step does not necessarily entail a training step; instead, a training step is carried out every
\texttt{train\_every} sampling steps. 
As backpropagation ~\cite{Rumelhart1986} on quantum devices is computationally intensive due to gradients being estimated via the parameter-shift rule~\cite{Mitarai2018, schuld2019evaluating}, we introduced this parameter as a means to keep the number of training steps per episode feasible. We note, however, that in this paper, we only report validation results on quantum hardware, while the agent has been trained in simulation. Similarly, we update the target network parameters to equal the policy network parameters every \texttt{update\_every} sampling steps. After the initial warm-up phase, we decay $\epsilon$ linearly  over \texttt{epsilon\_duration} sampling steps in total, starting at a value of \texttt{epsilon\_start=1}, and ending at a value of \texttt{epsilon\_end=0.01}.
Keeping $\epsilon > 0$ ensures continued exploration with a near-greedy policy. 

Since performance on the $\epsilon$-greedy policy is not indicative of learnt performance when $\epsilon$ 
is large~\cite{baker2016designing}, we estimate the expected return achieved by the current greedy policy in regular intervals. 
Specifically, we measure return over a single episode on a copy of the training environment every
\texttt{validate\_every=100} sampling steps (note that the parameter does not influence the actual
training process, and is just used for performance monitoring). If the average validation return over the past consecutive
25 validation steps reaches 196 (recall that the maximum return is 200, and that we need to allow
for some jitter), we regard the task as solved and terminate training early. While this
differs from the official \texttt{CartPole-v0} benchmark (see \url{https://gym.openai.com/envs/CartPole-v0/}) that necessitates a return of at least 195 sustained over 100 episodes, we find that training is very unlikely to
diverge past this point, given that $\epsilon$ has decayed sufficiently.\footnote{We provide a 
set of results on the accompanying \newline ~\website{} that have enjoyed traversing the maximum number
of episodes, and none of the results shows difference in convergence behaviour depending
on the convergence criterion used. However, for experiments on the experimental IBM Quantum device,
a reduced number of episodes is crucial to ensure practical feasibility of the 
calculations.}

\subsection{Encoding and Extraction Methods}
\label{sec:encoding-extraction}
After experimentally verifying the correctness of our implementation, we replace the Q-network by a VQC using the circuit architectures proposed in Refs.~\cite{Lockwood2020,skolik2021}. The need for mapping input parameters onto quantum states
has already been discussed in Sec.~\ref{sec:input-enc}; we consider the following approaches:
\begin{enumerate*}
    \item \textbf{Continuous (C)}: continuous encoding applied to all input components.
    \item \textbf{Scaled \& Continuous (SC)}: scaled encoding applied to finite-domain input components, continuous encoding otherwise.
    \item \textbf{Scaled \& Directional (SD)}: scaled encoding applied to finite-domain input components, directional encoding otherwise.
\end{enumerate*}
Along with the encoding strategies, we also investigate the impact of different Q-value extraction methods on agent performance. This is necessary due to the mismatch between VQC outputs and Q-values. In particular, we distinguish between:
\begin{enumerate*}
    \item \textbf{Local Scaling}: each output is scaled by a dedicated trainable weight as described in Ref.~\cite{skolik2021}.
    \item \textbf{Global Scaling (GS)}: all outputs are scaled by a single trainable weight.
    \item \textbf{Global Scaling with Quantum Pooling (GSP)}: quantum pooling as described in Ref.~\cite{Lockwood2020}, followed by global scaling.
\end{enumerate*}

\subsubsection{Initial Experiment}

We conducted experiments for each combination of input encoding, Q-value extraction method and circuit architecture,
totalling in \numprint{18} runs. To this end, we adapted hyperparameters from Ref.~\cite{skolik2021} to our slightly modified
algorithm described in Section~\ref{sec:methodology} (without data re-uploading). VQC
weights are initialised to zero 
to avoid \emph{barren plateaus}~\cite{mcclean_barren_2018}, i.e. the vanishing gradient
problem as suggested in Ref.~\cite{skolik2021layerwise} and classical weights are initialised
to one.

\begin{figure*}[htbp]
    \centering
    \vspace{-3em}
    \vspace*{0em}\input{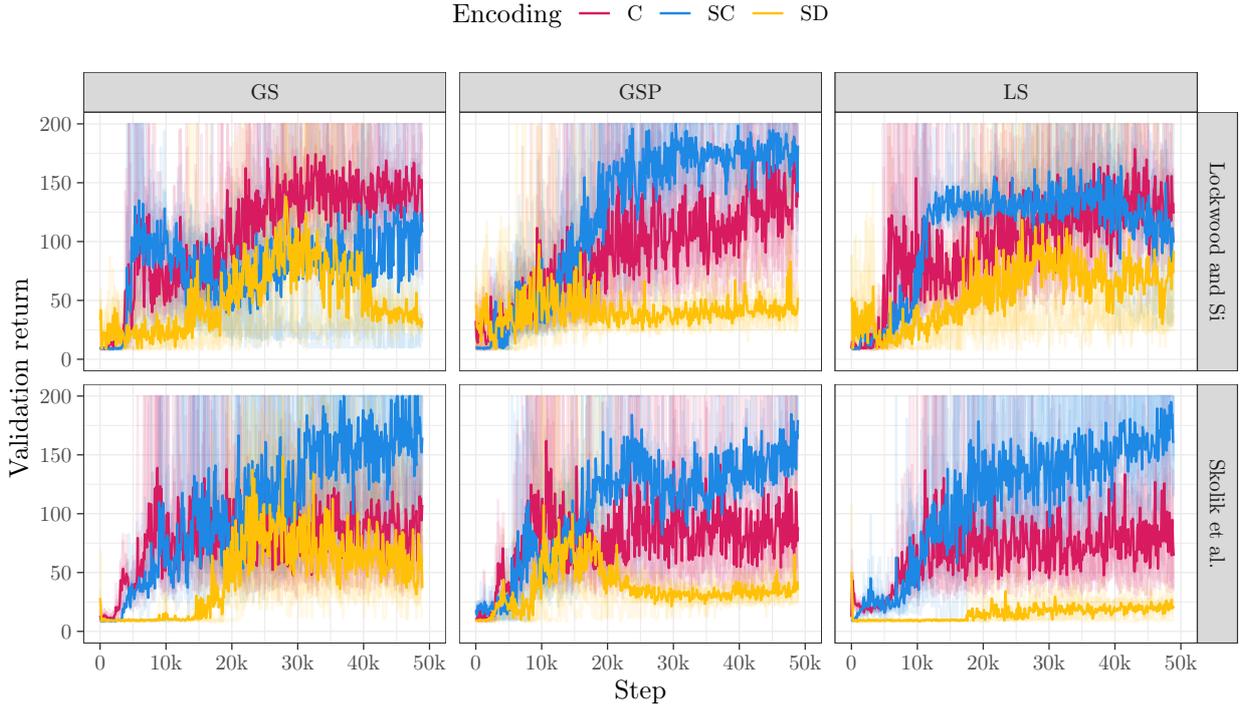}\vspace*{-1.7em}\vspace*{-1em}
    \caption{Validation returns for using the VQC-layer structure as in Ref.~\cite{Lockwood2020} (top) and Ref.~\cite{skolik2021} (bottom) with different input encoding
    strategies. The columns correspond to the extraction strategy (ltr. Global Scaling (GS), Global Scaling with Quantum Pooling (GSP), Local Scaling (LS), Ref. section~\ref{sec:encoding-extraction}).
    Results are averaged over five experiments each. These experiments are based on the \texttt{CartPole-v0} environment, where the maximum achievable return value is \numprint{200}.}
    \label{fig:skolik_hyper}
\end{figure*}
Results are shown in Fig.~\ref{fig:skolik_hyper}. As is apparent,
instabilities occur in every run and are not tied to a specific encoding-/extraction setting.
Nevertheless, some models only achieve comparatively low returns on average: In particular,
runs involving directional encoding tend to perform sub-par, which we attribute to the high
information-loss incurred by the encoding scheme. Directional encoding is therefore not
considered in further experiments.

To minimize the number of classical parameters, we focus on global scaling (with and without pooling) in further
experiments. While local scaling has not performed worse or less stable, the additional classical parameters
increase model capacity, and might therefore shadow deficiencies on the quantum parts.

As described by Mnih et al.~\cite{Mnih16}, Q-Learning is known to be instable, when a
nonlinear function approximator, such as a classical neural network or a VQ-DQN, is used to
represent the state-action value function. Our implementation already incorporates mechanisms
suggested by Mnih et al. to support convergence in Q-Learning.
However, these mechanisms do not guarantee a stable behaviour and we can not rule out that the
instabilities in VQ-DQN are caused by classical algorithmic constraints. In
section~\ref{sec:comparison} we compare the VQ-DQN to a classical neural network using the
same algorithm. Our results support the hypothesis that the reason for instabilities could be
classical. Therefore, in the next subsection, we study the effect of classical hyperparameters on the training process of VQ-DQN.

\subsection{Cross-Validation}
\label{sec:cross-validation}
As instabilities persist throughout our experiments, we turn to hyperparameters as a source of instabilities.
To this end, we re-utilize the above setting (C, SC/GS, GSP) with hyperparameters from Ref.~\cite{skolik2021}
as a starting point. Following recommendations~\cite{he2016, chen2020, brown2020, vaswani2017} from classical supervised learning, we add a
linear decay to the learning rate $\eta$. In particular, we decrease $\eta$ over a period of \texttt{eta\_duration}
training steps from \texttt{eta\_start} towards a target value of \texttt{eta\_end=0.01*eta\_start}. Additionally, we
progressively increase the \texttt{update\_every} parameters as learning progresses. This choice is motivated
by the observation that the delta between target and policy network decreases as the agent becomes more
proficient on the task. Finally, to optimize resource utilization
and minimize training time, we increase the batch size from 16 to 32, since this does not have a major impact on the agent's performance \cite{Bengio2012}. 

We cross-validate over the following hyper-parameter choices: \texttt{eta\_start} (i.e., the initial learning
rate) $\in \{10^{-3}, 10^{-2}, 10^{-1}\}$, \texttt{eta\_duration} (learning rate decay duration) $\in \{2000, 4000\}$,
\texttt{epsilon\_duration} $\in \{10000, 20000, 30000\}$, \texttt{gamma} $\in \{0.99, 0.999\}$. The remaining
parameters have been kept fixed over all experiments and are listed in Tab.~\ref{tab:hyperparameters}. The
following subsections describe our results obtained on the baseline setting (without data re-uploading),
and a modified variant with data re-uploading, respectively.

\begin{table*}[htbp]
    \caption{Hyperparameter settings for cross-validation.}\label{tab:hyperparameters}
    
    \definecolor{lightgray}{gray}{0.95}
    \rowcolors{3}{lightgray}{}
    \begin{tabularx}{\linewidth}{lXr}
        \toprule
         \rowcolor{white}Hyperparameter & Description & Default value\\\midrule
         \rowcolor{white}\multicolumn{3}{c}{\emph{Fixed parameters throughout cross validation runs}}\\
         \verb|num_steps| & \#sampling steps & \numprint{50000}\\
         \verb|train_after| & \#sampling steps before first training step & \numprint{1000}\\
         \verb|train_every| &\#sampling steps between training steps & 10\\
         \texttt{update\_every\_start} & initial \#sampling steps between target network updates & 30 \\
         \texttt{update\_every\_end} & final \#sampling steps between target network updates & 500 \\
         \texttt{update\_every\_duration} & \#sampling steps for \texttt{update\_every} increase & \numprint{35000} \\
         \verb|replay_capacity| & max. \#transitions in replay buffer & \numprint{50000}\\
         \verb|optimizer|  & Loss-function optimizer & Adam~\cite{kingma2014}\\
         \verb|batch_size| & batch size for gradient descent & 32 \\
         \verb|loss| & TD error loss function & \(L_{2}\) \\
         \verb|epsilon_start| & initial value for $\epsilon$ decay & 1.0 \\
         \verb|epsilon_end| & final value for $\epsilon$ decay & 0.01 \\
         \verb|validate_every| & \#sampling steps between validation runs & 100 \\
         \texttt{eta\_end} & final value for learning rate $\eta$ & $0.01*\texttt{eta\_start}$\\
         \cmidrule{1-3}
         \rowcolor{white}\multicolumn{3}{c}{\emph{Hyperparameters subject to cross validation}}\\
        \texttt{eta\_start} & initial value for learning rate $\eta$ & \{0.001, 0.01, 0.1\} \\
         \texttt{eta\_duration} & \#training steps for learning rate decay & \{\numprint{2000}, \numprint{4000}\} \\
         \texttt{epsilon\_duration} & \#sampling steps for $\epsilon$ decay & \{\numprint{10000}, \numprint{20000}, \numprint{30000}\} \\
         \texttt{gamma} & discount factor $\gamma$ & \{0.99, 0.999\} \\
         \bottomrule
    \end{tabularx}
\end{table*}

\subsubsection{Baseline}
\label{sec:baseline}
\begin{figure*}[]
    \centering
    \vspace*{0em}\input{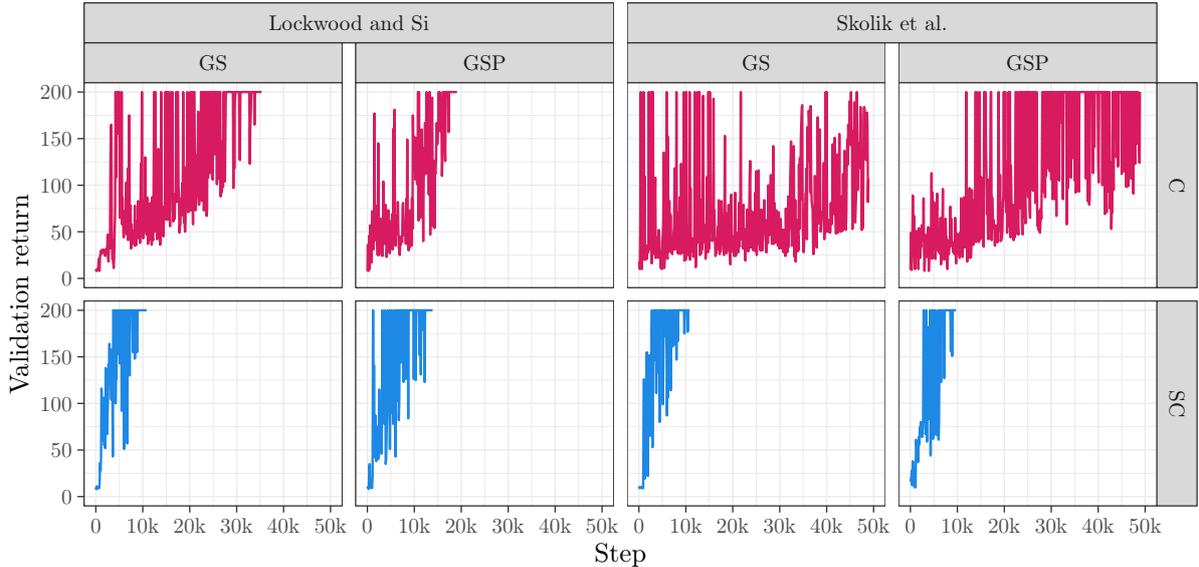}\vspace*{-1.7em}
    \caption{Validation returns for our best-performing hyperparameter constellations in the baseline configurations. The figure considers the results for the VQC architecture described in Ref.~\cite{Lockwood2020} (left) and Ref.~\cite{skolik2021} (right). Columns correspond to the extraction method (ltr. Global Scaling (GS), Global Scaling with Quantum Pooling (GSP), Local Scaling (LS)), rows correspond to the input encoding strategy (Continuous (C), Scaled \& Continuous (SC), see Section~\ref{sec:encoding-extraction}). The experiments are based on the \texttt{CartPole-v0} environment, where the maximum achievable return value is \numprint{200}.}
    \label{fig:baseline_xval}
\end{figure*}

Results for the baseline case are depicted in Fig.~\ref{fig:baseline_xval} and Tab.~\ref{tab:hyperparameters_results}. We only present a 
selection of the best-performing hyperparameter constellations due to space constraints, but
provide the full set of results on the accompanying~\website. As evident from the figure,
almost every model was able to achieve stable optimal performance (according to our early-stopping criterion). 
Generally, the SC encoding tends to convergence faster as compared to models with continuous encoding; in 
the best case (Skolik et al./SC/GSP), optimal performance is reached after a mere 97 validation
steps. This shows that VQ-DQN is in fact capable of learning a stable optimal policy, albeit
hyperparameter tuning is a sensitive influence factor.

\subsubsection{Baseline with data re-uploading}
\label{sec:baseline_with_data_re}

\begin{figure*}[htbp]
    \centering
    \vspace*{-2em}\vspace*{0em}\input{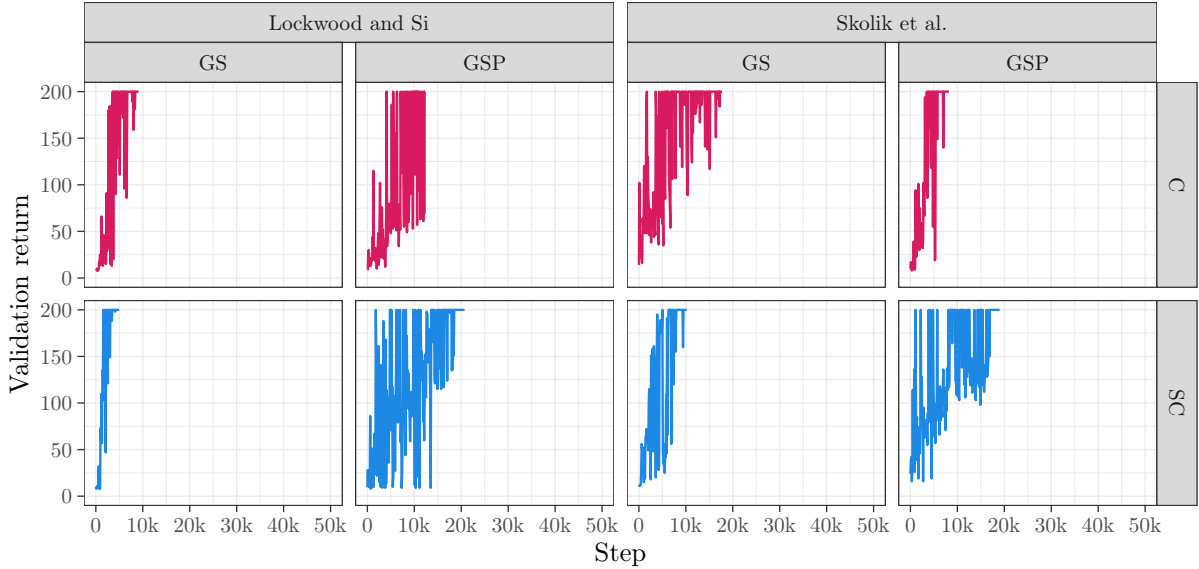}\vspace*{-1.7em}\vspace*{-1em}
    \caption{Validation returns for our best-performing hyperparameter constellations in the baseline configurations with data re-uploading. The figure considers the results for the VQC architecture described in Ref.~\cite{Lockwood2020} (left) and Ref.~\cite{skolik2021} (right). Columns correspond to the extraction method (ltr. Global Scaling (GS), Global Scaling with Quantum Pooling (GSP), Local Scaling (LS)), rows correspond to the input encoding strategy (Continuous (C), Scaled \& Continuous (SC), see Section~\ref{sec:encoding-extraction}). The experiments are based on the \texttt{CartPole-v0} environment, where the maximum achievable return value is \numprint{200}.}
    \label{fig:baseline_withDataRe_xval}
\end{figure*}

From Fig.~\ref{fig:baseline_xval}, it is evident that the performance of the VQ-DQN algorithm also suffers due to the choice of encoding strategy used along with the bad choice of hyperparameters. For example, the agent with the continuous encoding format does not learn an optimal policy in many cases. Here to increase the expressivity of the model, we can use techniques such as data re-uploading ~\cite{skolik2021, salinas2020}.  The results for the baseline case with data re-uploading are depicted in Fig.~\ref{fig:baseline_withDataRe_xval} and Tab.~\ref{tab:hyperparameters_results}. As in Sec.~\ref{sec:baseline}, We only present a selection of the best-performing hyperparameter constellations due to space constraints.
From the results shown in Fig.~\ref{fig:baseline_withDataRe_xval}, we can conclude that the data re-uploading strategy does not significantly increase the VQ-DQN algorithm's performance. Though it increases the expressive power of the model, which in turn allows the agent to learn optimal behavior in some cases (for example, agent with Continuous (C) encoding), the performance change is negligible or even negative in most cases. Moreover, the data re-uploading strategy increases the gate count in the VQC architecture, and this increase in gate count is not ideal for the NISQ devices due to noise.

\begin{table*}[]
\newcommand{\na}{\multicolumn{1}{c}{-}}
\newcommand{\arch}[3]{#1/#2/#3}
\caption{Hyperparameters cross-validation results. The table provides values for \texttt{eta\_start} (\(\eta_{\text{s}}\)),
\texttt{eta\_duration} (\(\eta_{\text{d}}\)),
\texttt{epsilon\_duration} (\(\epsilon_{\text{s}}\)), and
\texttt{gamma} (\(\gamma\)). Encodings C, SC, GS, and GSP as defined in Section~\ref{sec:encoding-extraction}.}\label{tab:hyperparameters_results}

\definecolor{lightgray}{gray}{0.95}
\rowcolors{3}{lightgray}{}
\centering\begin{tabular}{l@{\hspace{2em}}S[table-format=1.3]rrS[table-format=1.3]@{\hspace{2em}}S[table-format=1.2]rrS[table-format=1.3]}
\toprule
\multirow{2}{*}{Architecture} & \multicolumn{4}{c}{Baseline} & \multicolumn{4}{c}{Baseline + data re-uploading} \\ 
 & \(\eta_{\text{s}}\) & \(\eta_{\text{d}}\) & \(\epsilon_{\text{d}}\) & \(\gamma\) & \(\eta_{\text{s}}\) & \(\eta_{\text{d}}\) & \(\epsilon_{\text{d}}\) & \(\gamma\) \\ 
\midrule
\arch{\cite{Lockwood2020}}{C}{GSP} & 0.01 & \numprint{2000} & \numprint{20000} & 0.99 & 0.01 & \numprint{2000} & \numprint{30000} & 0.99 \\
\arch{\cite{Lockwood2020}}{C}{GS} & 0.001 & \numprint{4000} & \numprint{20000} & 0.99 & 0.01 & \numprint{2000} & \numprint{30000} & 0.999 \\
\arch{\cite{Lockwood2020}}{SC}{GSP} & 0.01 & \numprint{2000} & \numprint{20000} & 0.99 & 0.1 & \numprint{2000} & \numprint{20000} & 0.999 \\
\arch{\cite{Lockwood2020}}{SC}{GS} & 0.01 & \numprint{4000} & \numprint{30000} & 0.99 & 0.01 & \numprint{2000} & \numprint{30000} & 0.99 \\
\arch{\cite{skolik2021}}{C}{GSP} & \na & \na & \na & \na & 0.01 & \numprint{2000} & \numprint{30000} & 0.999 \\
\arch{\cite{skolik2021}}{C}{GS} & \na & \na & \na & \na & 0.01 & \numprint{2000} & \numprint{10000} & 0.99 \\
\arch{\cite{skolik2021}}{SC}{GSP}& 0.01 & \numprint{2000} & \numprint{10000} & 0.999 & 0.01 & \numprint{2000} & \numprint{10000} & 0.99 \\
\arch{\cite{skolik2021}}{SC}{GS} & 0.01 & \numprint{4000} & \numprint{30000} & 0.99 & 0.01 & \numprint{2000} & \numprint{10000} & 0.99\\
\bottomrule
\end{tabular}
\end{table*}

\subsection{Discussion on Instabilities in VQ-DQN}
\label{sec:result_discussion}

In our approach of VQ-DQNs, we train a VQC with a classical optimization loop. Such a setting is known to be prone to the \emph{barren plateau} effect~\cite{mcclean_barren_2018}, which describes a problem of vanishing gradients that causes the inability to converge to an optimal return value. However, \emph{barren plateaus} only occur in random VQCs. To counter randomness in the quantum circuits, we initialised all VQC parameters systematically to zeros. Since the VQCs in our experiments are neither very wide (4 Qubits), nor deep (5 \enquote{Layers}), randomness induced by gradient-based optimization is also limited. Therefore we rule out \emph{barren plateaus} as the source of instabilities.

With the possibility of the barren plateau avoided, one can say that every agent with its unique architecture combinations and a reasonable encoding scheme is capable of learning the optimal policy to solve the cart pole environment. This can be seen from the results shown in sections \ref{sec:baseline} and \ref{sec:baseline_with_data_re}. The architecture combinations which did not learn an optimal policy during experiments conducted by different authors (Ref \cite{skolik2021} and \cite{Lockwood2020}) showed a tendency to learn the optimal policy during our experiments. The reason why these agents show such a tendency is the selection of the right set of classical hyperparameters. The agents learned the optimal policy only for a few sets of classical hyperparameters during our hyperparameters search. This made us conclude that the VQC-DQN algorithms are highly sensitive to classical hyperparameters.  \\
The results from sections \ref{sec:baseline} and \ref{sec:baseline_with_data_re} elucidate that the data re-uploading strategy does not always outperform its corresponding architecture without data re-uploading in sampling efficiency. One possible reason for this could be that the optimal hyperparameter set required for these architectures might fall outside the search space used in the experiments. One other possible reason for this underperformance can be inferred from the work of Schuld et al. \cite{Schuld2021}. Schuld et al. show that the function represented by a VQC is a Fourier sum. In particular, the variational layers determine the amplitudes and the encoding layers determine the frequency spectrum. As shown in ref \cite{Periyasamy2022}, when it comes to data re-uploading strategy, the variational layers between two encoding layers might not be expressive enough which reduces the overall expressivity of the VQC. The expressivity can be increased by increasing the number of variational layers. However, this leads to an architectural change which is out of scope for this study.

\section{Validation on IBM Quantum Device}
\label{sec:ibmq_validation}
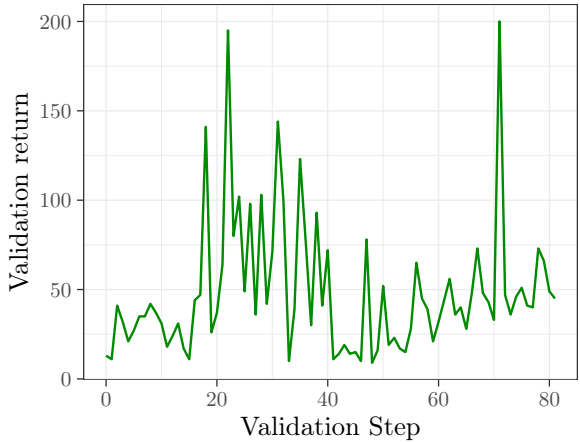
\begin{figure}[htbp]
    \centering
    \vspace{-2.5em}
    \vspace*{0em}\input{img/ibmq.tex}\vspace*{-1.7em}
    \vspace{-2.5em}
    \caption{Results of our validation run on \texttt{ibmq\_ehningen}~\cite{ibmq2021}. The experiments are based on the \texttt{CartPole-v0} environment, where the maximum achievable return value is \numprint{200}.}
    \label{fig:ibmq}
\end{figure}

Results from Sec.~\ref{sec:baseline} and Sec.~\ref{sec:baseline_with_data_re} illustrate that a VQC can learn a stable policy to solve the \texttt{CartPole-v0}
environment using the DQN algorithm if the right set of hyperparameters are used. In order to gauge the detrimental influence of device noise on an agent trained using an ideal simulator in solving the environment, we
tested the trained model in an actual IBM quantum device~\cite{ibmq2021}.  As a first step, we had to
port the VQ-DQN algorithm from the Tensorflow/TFQ API~\cite{Abadi15, Broughton2021}  to the Pytorch/Qiskit API~\cite{Abraham19, Paszke2019} as the IBM quantum devices use the Qiskit API~\cite{Abraham19} as their primary
programming library. There is one significant difference between the Qiskit API~\cite{Abraham19} and the TFQ API~\cite{Broughton2021} to be noted here. The TFQ~\cite{Broughton2021} API calculates the expectation value analytically, whereas the Qiskit API~\cite{Abraham19} estimates the expectation value by simulating the \texttt{ideal quantum device} and measuring its outcomes. Likewise, the expectation values are estimated in the IBM quantum device~\cite{ibmq2021} by measuring the outcome multiple times. Further, we trained the best-performing model without data re-uploading from Sec.~ \ref{sec:baseline} using Qiskit \texttt{qasm\_simulator}~\cite{Abraham19} and verified the
correctness of our implementation in comparison to the results from Sec.~\ref{sec:baseline}. We chose a model without data re-uploading due to the fact that the quantum devices available right now are prone to noise. Hence adding more gates via data re-uploading in NISQ devices seems counter-productive. 
Once the correctness was
verified, we uploaded the weights trained using the \texttt{qasm\_simulator} to the IBM Quantum  (ibmq\_ehningen) device and validated the learned policy. The results of these validation runs are
shown in Fig.~\ref{fig:ibmq}. 

Though the agents trained in the ideal simulator learned an optimal policy to solve the \texttt{Cartpole-v0} environment, testing the trained agent in the ibmq\_ehningen device did not reproduce the optimal behavior. This degradation in behavior is due to the noise present in the IBM Quantum device. An agent trained in the IBM Quantum device from scratch might reduce the effect of noise and learn a policy close to the optimal policy. Additionally, different types of error mitigation techniques can be employed to reduce the effects of noise at the cost of additional overhead.
However, when we attempted to train the agent from scratch on the IBM quantum device, the training turned out to be infeasible due to the following practical issues:

\begin{enumerate*}
    \item We observed waiting times in the queue to start a job execution (referred to as fair-share queue for jobs in IBM Quantum systems) in the cloud-based IBM Quantum device that were typically two orders of magnitude (or more) larger than the actual job execution time. As (roughly speaking) a single action selection corresponds to a single job in the fair share queue, even completion of a single episode takes a substantial amount of time.
    \item The overall time it takes to achieve low-variance estimators of expectation values can become quite large due to the large number of shots (i.e., measurement samples) taken for a single circuit instance.
\end{enumerate*}

Here, the first hindrance can be overcome in time as the availability of quantum devices and resources is expected to increase in the near future. As improvements in hardware and orchestration of quantum and classical computational resources progress, we might also be witness to an increased number of circuit layer operations per second (CLOPS)~\cite{Andrew2021}. When we started the training process in the \texttt{ibmq\_ehningen} device, the job execution time for each action selection took between 15 to 30 seconds, and each training step took around 3 minutes (as the training step performs gradient decent via parameter-shift rule). These long execution and waiting times make the training process in real quantum devices impractical for training algorithms like VQ-DQN, where the agent has to interact with the environment sequentially.

\section{Comparison to classical Neural Network}
\label{sec:comparison}

A popular ``quantum advantage'' claimed by a good fraction of the literature in QRL is that the VQC has better state-action pair representation, samples efficiently, and learns an optimal policy faster than the classical neural network~\cite{vqdqn, Lockwood2020, skolik2021}.
Hence to compare the sample efficiency of a VQ-DQN-agent trained on an ideal simulator against a classical neural network, we trained a simple fully-connected network with one hidden layer to solve the \texttt{Cartpole-v0} environment. To ensure a fair comparison, we restricted the total number of parameters of the network to 58,
and did cross-validation on the same set of hyperparameters as explained in Sec.~\ref{sec:cross-validation}. 

\begin{figure}[htbp]
    \centering
    \vspace*{0em}\input{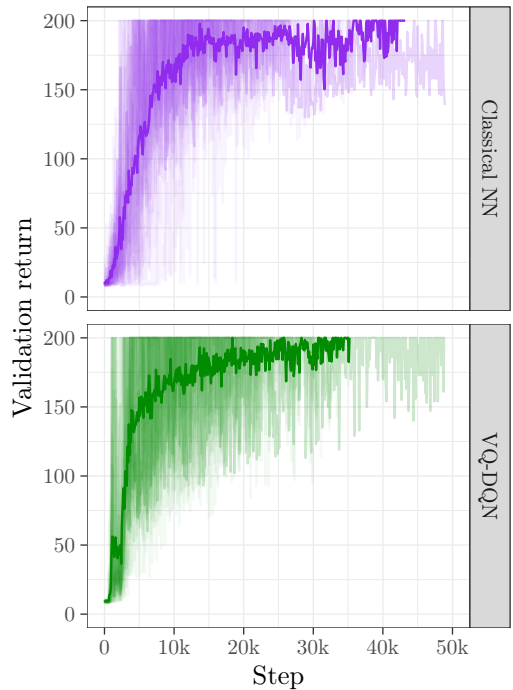}\vspace*{-1.7em}
    \vspace{-1em}
    \caption{Comparison between VQ-DQN and classical NN averaged over 30 different agents. The experiments are based on the \texttt{CartPole-v0} environment, where the maximum achievable return value is \numprint{200}.}
    \label{fig:stat_testing}
\end{figure}

The results shown in Fig.~\ref{fig:stat_testing} indicate that initially, the VQC seems to learn faster than the neural network. 
For a more rigorous discussion we resort to Ref.~\cite{Kakade2003}, where sample efficiency of an algorithm is defined for an online learning setting as the
number of time steps from which on an agent trained by the algorithm perceives an average reward exceeding
a certain threshold $V_\mathrm{thresh}$ with high probability.

For a weaker statement adapted to a numerical treatment, we propose to use significance testing under the null
hypothesis of mean reward being smaller than $V_\mathrm{thresh}$.
Thus, we define sample efficiency as the number of time steps from which on the null hypothesis is rejected with 
respect to the given threshold.
As statistical test we propose to use a one-sample $t$-test \cite{Student,Henderson2017}, in particular its one-sided version as we compare the performance of a particular algorithm against a given threshold. Thus, we perform sufficiently many independent runs of each algorithm  and fix the significance level at $\alpha=0.05$.

With respect to this metric the variational quantum circuit indeed crosses $V_\mathrm{thresh}=120$
faster than the classical network; however, for larger threshold values, no definite statement can be made.

\section{Conclusion}\label{sec:conclusion}
We have systematically studied the performance of quantum-assisted 
reinforcement learning schemes on both simulators and physical quantum computers.
We find---not quite unexpected---that at the current, early state of technological
development, quantum computers do not bring any measurable advantage in this scenario.
We even find that simulated quantum systems do not bring clear advantages over
classical approaches.

Nonetheless, a number of constructive insights can be drawn from our experiments. 
Following previous work, we have trained models on classical simulators and only
performed the execution step on quantum hardware. This approach, albeit practically
necessitated by current-day hardware, creates a mis-match in terms of handling noise:
For future work, we recommend including noise in the training process,
especially since Ref.~\cite{Meyer2021} suggests for small-scale systems that
existing noise models lead to a good match between simulation and hardware,
and therefore provide a more faithful basis for comparing between algorithmic performance
on simulated and physical hardware.

Most importantly, our results do \emph{not} corroborate observations made when reinforcement
learning on quantum computers was first introduced into the literature in Ref.~\cite{vqdqn}: 
While the authors in this approach upload weights determined by classical training onto a quantum machine
as we do in this paper, they  find that executing the model does not vary much between simulation
and NISQ machine. We, on the contrary, observe a total mismatch in performance. 
We expect the most probable explanation for this discrepancy to lie in (a) the size of the machine
(five versus 27 qbits) and the problem of choice (cognitive-radio versus cartpole; a random policy
as would be caused by growing amounts of noise from NISQ devices is obviously better suited
to the former than the latter). 

We encounter additional hindrances towards the practical application of quantum computers:
Waiting time on queues in a shared, cloud-like environment is a major practical issue,
which will however be alleviated with the broader availability of quantum chips. Nonetheless,
the temporal contributions of sequential elements of algorithms to the overall computation
time would also occur in a non-shared setting and do substantially increase wall-time
run-times, which is an obvious impediment to practical utility.

As long as noise and imperfections are unavoidable, we find that adapting algorithms
and approaches to account for these issues is a major design challenge for quantum algorithms.
One possible approach would be to equip simulated QPU designs with appropriate, yet tunable and
physically realistic noise behaviour. By seeking optimal models and parameters
under these unavoidable constraints, an ``ideal'' noise model can be identified, and
future QPUs be built such that design trade-off decisions are taken so that the resulting
hardware closely mimics the identified noise and imperfection behaviour.
In other words, we hypothesise that in the space of hardware design decisions, and
assuming that hardware imperfections impact different computations in a different way,
this opens a degree of freedom that can be leveraged to design custom algorithmic-specific
hardware.

\section*{Acknowledgements} 
We acknowledge the use of IBM Quantum services for this work. The views expressed are those of the authors, and do not reflect the official policy or position of IBM or the IBM Quantum team.

\textbf{Funding}: This work was supported by the German Federal Ministry of Education and Research (BMBF),
funding program ``quantum technologies -- from basic research to market'', grant number 13N15645.

\bibliographystyle{elsarticle-num} 
\bibliography{bibliography}

\appendix
\section{Reparameterization}
\label{sec:appendix_a}

As part of our initial investigation into potential sources of divergent behavior, we analysed model weight distributions over time. Here, we noticed that in some runs, the magnitudes of the learned VQC weights increase indefinitely over time. This behavior is intriguing, considering that the respective parameters control qubit rotations (in X-, Y- or Z-basis) in radians, and therefore naturally ``wrap'' at $2\pi$ (in the sense that a parameter $\theta$ and $\theta + 2\pi$ specify the same circuit). The periodicity in parameter space translates to a periodicity in the loss function. This can intuitively be thought of as ``copies'' of the loss landscape at $2\pi$ increments along the respective dimensions. In consequence, circuit weights growing beyond $2\pi$ imply that an adjacent copy of the optimization landscape is visited and a potential minimizing value along this dimension has been overshot.

To rule out that this phenomenon impedes learning, we modify the quantum circuit by squashing the (unbounded) VQC parameters $\theta$ to the range $[0;2\pi]$ using the transformation $f(\theta) = 2 \pi \cdot \sigma(\theta)$ (where $\sigma$ is the well-known sigmoid function). This effectively ensures that any given set of parameter values uniquely specifies the qubit transformation enacted by the circuit, while being optimizable by backpropagation.

Unfortunately, all of our experiments involving reparameterizations did not yield better performance or further insights. We take this as evidence that periodicities are, after all, not the root cause of instabilities in VQ-DQN.

\end{document}

%% file: img/vqc_standard.tex
\scalebox{0.7}{%
\begin{quantikz}[column sep=2.5pt, row sep=3pt, thin lines]
\lstick{\ket{0}} & \qw & \gate{R_x(x_1)}\gategroup[wires=4,steps=1,style={inner sep=0pt, dotted}]{\scriptsize{encoding layer}} & \qw & \qw & \qw & \gate{R_y}\gategroup[wires=4,steps=6,style={inner sep=0pt, dashed}]{\scriptsize{variational layer}} & \gate{R_z} & \ctrl{1}  & \qw & \qw & \control{} & \qw & \qw & \qw & \gate{R_y}\gategroup[wires=4,steps=6,style={inner sep=0pt, dashed}]{} & \gate{R_z} & \ctrl{1} & \qw & \qw & \control{} & \qw & \qw & \qw & \qw\gategroup[wires=4,steps=3,style={inner sep=0pt, dashed}]{} & \ldots &  & \qw & \qw & \qw \\
\lstick{\ket{0}} & \qw & \gate{R_x(x_2)} & \qw & \qw & \qw & \gate{R_y} & \gate{R_z} & \control{} & \ctrl{1} & \qw & \qw & \qw & \qw & \qw & \gate{R_y} & \gate{R_z} & \control{} & \ctrl{1} & \qw & \qw & \qw & \qw & \qw & \qw & \ldots & & \qw & \qw & \qw \\
\lstick{\ket{0}} & \qw & \gate{R_x(x_3)} & \qw & \qw & \qw & \gate{R_y} & \gate{R_z} & \qw & \control{} & \ctrl{1} & \qw & \qw & \qw & \qw & \gate{R_y} & \gate{R_z} & \qw & \control{} & \ctrl{1} & \qw & \qw & \qw & \qw & \qw & \ldots & & \qw & \qw & \qw \\
\lstick{\ket{0}} & \qw & \gate{R_x(x_4)} & \qw & \qw & \qw & \gate{R_y} & \gate{R_z} & \qw & \qw & \control{} & \ctrl{-3} & \qw & \qw & \qw & \gate{R_y} & \gate{R_z} & \qw & \qw & \control{} & \ctrl{-3} & \qw & \qw & \qw & \qw & \ldots & & \qw & \qw & \qw 
\end{quantikz}%
}%

%% file: img/vqc_data_reuploading.tex
\hspace*{-0.7em}%
\scalebox{0.7}{%
\begin{quantikz}[column sep=2.5pt, row sep=3pt, thin lines]
\lstick{\ket{0}} & \qw & \gate{R_x(x_1)}\gategroup[wires=4,steps=1,style={inner sep=0pt, dotted}]{\scriptsize{encoding layer}} & \qw & \qw & \qw & \gate{R_y}\gategroup[wires=4,steps=6,style={inner sep=0pt, dashed}]{\scriptsize{variational layer}} & \gate{R_z} & \ctrl{1}  & \qw & \qw & \control{} & \qw & \qw & \qw & \gate{R_x(x_1)}\gategroup[wires=4,steps=1,style={inner sep=0pt, dotted}]{} & \qw & \qw & \qw & \gate{R_y}\gategroup[wires=4,steps=6,style={inner sep=0pt, dashed}]{} & \gate{R_z} & \ctrl{1} & \qw & \qw & \control{} & \qw & \qw & \qw & \qw\gategroup[wires=4,steps=3,style={inner sep=0pt, dotted}]{} & \ldots & & \qw & \qw & \qw & \qw\gategroup[wires=4,steps=3,style={inner sep=0pt, dashed}]{} & \ldots & & \qw & \qw & \qw \\
\lstick{\ket{0}} & \qw & \gate{R_x(x_2)} & \qw & \qw & \qw & \gate{R_y} & \gate{R_z} & \control{} & \ctrl{1} & \qw & \qw & \qw & \qw & \qw & \gate{R_x(x_2)} & \qw & \qw & \qw & \gate{R_y} & \gate{R_z} & \control{} & \ctrl{1} & \qw & \qw & \qw & \qw & \qw & \qw & \ldots & & \qw & \qw & \qw & \qw & \ldots & & \qw & \qw & \qw \\
\lstick{\ket{0}} & \qw & \gate{R_x(x_3)} & \qw & \qw & \qw & \gate{R_y} & \gate{R_z} & \qw & \control{} & \ctrl{1} & \qw & \qw & \qw & \qw & \gate{R_x(x_3)} & \qw & \qw & \qw & \gate{R_y} & \gate{R_z} & \qw & \control{} & \ctrl{1} & \qw & \qw & \qw & \qw & \qw & \ldots & & \qw & \qw & \qw & \qw & \ldots & & \qw & \qw & \qw \\
\lstick{\ket{0}} & \qw & \gate{R_x(x_4)} & \qw & \qw & \qw & \gate{R_y} & \gate{R_z} & \qw & \qw & \control{} & \ctrl{-3} & \qw & \qw & \qw & \gate{R_x(x_4)} & \qw & \qw & \qw & \gate{R_y} & \gate{R_z} & \qw & \qw & \control{} & \ctrl{-3} & \qw & \qw & \qw & \qw & \ldots & & \qw & \qw & \qw & \qw & \ldots & & \qw & \qw & \qw 
\end{quantikz}%
}%

%% file: img/ibmq.tex
\begin{tikzpicture}[x=1pt,y=1pt]
\definecolor{fillColor}{RGB}{255,255,255}
\path[use as bounding box,fill=fillColor,fill opacity=0.00] (0,0) rectangle (222.67,222.67);
\begin{scope}
\path[clip] (  0.00, 24.06) rectangle (222.67,198.61);
\definecolor{drawColor}{RGB}{255,255,255}
\definecolor{fillColor}{RGB}{255,255,255}

\path[draw=drawColor,line width= 0.5pt,line join=round,line cap=round,fill=fillColor] (  0.00, 24.06) rectangle (222.67,198.61);
\end{scope}
\begin{scope}
\path[clip] ( 32.83, 51.96) rectangle (217.67,193.61);
\definecolor{fillColor}{RGB}{255,255,255}

\path[fill=fillColor] ( 32.83, 51.96) rectangle (217.67,193.61);
\definecolor{drawColor}{gray}{0.92}

\path[draw=drawColor,line width= 0.3pt,line join=round] ( 32.83, 69.18) --
	(217.67, 69.18);

\path[draw=drawColor,line width= 0.3pt,line join=round] ( 32.83,102.89) --
	(217.67,102.89);

\path[draw=drawColor,line width= 0.3pt,line join=round] ( 32.83,136.60) --
	(217.67,136.60);

\path[draw=drawColor,line width= 0.3pt,line join=round] ( 32.83,170.32) --
	(217.67,170.32);

\path[draw=drawColor,line width= 0.3pt,line join=round] ( 61.98, 51.96) --
	( 61.98,193.61);

\path[draw=drawColor,line width= 0.3pt,line join=round] (103.47, 51.96) --
	(103.47,193.61);

\path[draw=drawColor,line width= 0.3pt,line join=round] (144.96, 51.96) --
	(144.96,193.61);

\path[draw=drawColor,line width= 0.3pt,line join=round] (186.45, 51.96) --
	(186.45,193.61);

\path[draw=drawColor,line width= 0.5pt,line join=round] ( 32.83, 52.33) --
	(217.67, 52.33);

\path[draw=drawColor,line width= 0.5pt,line join=round] ( 32.83, 86.04) --
	(217.67, 86.04);

\path[draw=drawColor,line width= 0.5pt,line join=round] ( 32.83,119.75) --
	(217.67,119.75);

\path[draw=drawColor,line width= 0.5pt,line join=round] ( 32.83,153.46) --
	(217.67,153.46);

\path[draw=drawColor,line width= 0.5pt,line join=round] ( 32.83,187.17) --
	(217.67,187.17);

\path[draw=drawColor,line width= 0.5pt,line join=round] ( 41.23, 51.96) --
	( 41.23,193.61);

\path[draw=drawColor,line width= 0.5pt,line join=round] ( 82.72, 51.96) --
	( 82.72,193.61);

\path[draw=drawColor,line width= 0.5pt,line join=round] (124.21, 51.96) --
	(124.21,193.61);

\path[draw=drawColor,line width= 0.5pt,line join=round] (165.70, 51.96) --
	(165.70,193.61);

\path[draw=drawColor,line width= 0.5pt,line join=round] (207.19, 51.96) --
	(207.19,193.61);
\definecolor{drawColor}{RGB}{0,139,0}

\path[draw=drawColor,line width= 0.9pt,line join=round] ( 41.23, 61.09) --
	( 43.30, 59.74) --
	( 45.38, 79.97) --
	( 47.45, 73.90) --
	( 49.53, 66.49) --
	( 51.60, 70.53) --
	( 53.68, 75.92) --
	( 55.75, 75.92) --
	( 57.83, 80.64) --
	( 59.90, 77.27) --
	( 61.98, 73.23) --
	( 64.05, 64.46) --
	( 66.12, 68.51) --
	( 68.20, 73.23) --
	( 70.27, 63.79) --
	( 72.35, 59.74) --
	( 74.42, 81.99) --
	( 76.50, 84.01) --
	( 78.57,147.39) --
	( 80.65, 69.86) --
	( 82.72, 77.27) --
	( 84.80, 95.48) --
	( 86.87,183.80) --
	( 88.94,106.26) --
	( 91.02,121.10) --
	( 93.09, 85.36) --
	( 95.17,118.40) --
	( 97.24, 76.60) --
	( 99.32,121.77) --
	(101.39, 80.64) --
	(103.47,100.20) --
	(105.54,149.41) --
	(107.62,119.75) --
	(109.69, 59.07) --
	(111.76, 78.62) --
	(113.84,135.26) --
	(115.91,104.92) --
	(117.99, 72.55) --
	(120.06,115.03) --
	(122.14, 79.97) --
	(124.21,100.87) --
	(126.29, 59.74) --
	(128.36, 61.77) --
	(130.44, 65.14) --
	(132.51, 61.77) --
	(134.58, 62.44) --
	(136.66, 59.07) --
	(138.73,104.92) --
	(140.81, 58.39) --
	(142.88, 63.11) --
	(144.96, 87.39) --
	(147.03, 65.14) --
	(149.11, 67.83) --
	(151.18, 63.79) --
	(153.26, 62.44) --
	(155.33, 71.20) --
	(157.40, 96.15) --
	(159.48, 82.67) --
	(161.55, 78.62) --
	(163.63, 66.49) --
	(165.70, 73.90) --
	(167.78, 81.99) --
	(169.85, 90.08) --
	(171.93, 76.60) --
	(174.00, 79.30) --
	(176.08, 71.20) --
	(178.15, 84.69) --
	(180.22,101.54) --
	(182.30, 84.69) --
	(184.37, 81.32) --
	(186.45, 74.58) --
	(188.52,187.17) --
	(190.60, 84.01) --
	(192.67, 76.60) --
	(194.75, 83.34) --
	(196.82, 86.71) --
	(198.90, 79.97) --
	(200.97, 79.30) --
	(203.04,101.54) --
	(205.12, 96.83) --
	(207.19, 85.36) --
	(209.27, 82.67);
\definecolor{drawColor}{gray}{0.20}

\path[draw=drawColor,line width= 0.5pt,line join=round,line cap=round] ( 32.83, 51.96) rectangle (217.67,193.61);
\end{scope}
\begin{scope}
\path[clip] (  0.00,  0.00) rectangle (222.67,222.67);
\definecolor{drawColor}{gray}{0.30}

\node[text=drawColor,anchor=base east,inner sep=0pt, outer sep=0pt, scale=  0.80] at ( 28.33, 49.57) {0};

\node[text=drawColor,anchor=base east,inner sep=0pt, outer sep=0pt, scale=  0.80] at ( 28.33, 83.28) {50};

\node[text=drawColor,anchor=base east,inner sep=0pt, outer sep=0pt, scale=  0.80] at ( 28.33,116.99) {100};

\node[text=drawColor,anchor=base east,inner sep=0pt, outer sep=0pt, scale=  0.80] at ( 28.33,150.71) {150};

\node[text=drawColor,anchor=base east,inner sep=0pt, outer sep=0pt, scale=  0.80] at ( 28.33,184.42) {200};
\end{scope}
\begin{scope}
\path[clip] (  0.00,  0.00) rectangle (222.67,222.67);
\definecolor{drawColor}{gray}{0.20}

\path[draw=drawColor,line width= 0.5pt,line join=round] ( 30.33, 52.33) --
	( 32.83, 52.33);

\path[draw=drawColor,line width= 0.5pt,line join=round] ( 30.33, 86.04) --
	( 32.83, 86.04);

\path[draw=drawColor,line width= 0.5pt,line join=round] ( 30.33,119.75) --
	( 32.83,119.75);

\path[draw=drawColor,line width= 0.5pt,line join=round] ( 30.33,153.46) --
	( 32.83,153.46);

\path[draw=drawColor,line width= 0.5pt,line join=round] ( 30.33,187.17) --
	( 32.83,187.17);
\end{scope}
\begin{scope}
\path[clip] (  0.00,  0.00) rectangle (222.67,222.67);
\definecolor{drawColor}{gray}{0.20}

\path[draw=drawColor,line width= 0.5pt,line join=round] ( 41.23, 49.46) --
	( 41.23, 51.96);

\path[draw=drawColor,line width= 0.5pt,line join=round] ( 82.72, 49.46) --
	( 82.72, 51.96);

\path[draw=drawColor,line width= 0.5pt,line join=round] (124.21, 49.46) --
	(124.21, 51.96);

\path[draw=drawColor,line width= 0.5pt,line join=round] (165.70, 49.46) --
	(165.70, 51.96);

\path[draw=drawColor,line width= 0.5pt,line join=round] (207.19, 49.46) --
	(207.19, 51.96);
\end{scope}
\begin{scope}
\path[clip] (  0.00,  0.00) rectangle (222.67,222.67);
\definecolor{drawColor}{gray}{0.30}

\node[text=drawColor,anchor=base,inner sep=0pt, outer sep=0pt, scale=  0.80] at ( 41.23, 41.95) {0};

\node[text=drawColor,anchor=base,inner sep=0pt, outer sep=0pt, scale=  0.80] at ( 82.72, 41.95) {20};

\node[text=drawColor,anchor=base,inner sep=0pt, outer sep=0pt, scale=  0.80] at (124.21, 41.95) {40};

\node[text=drawColor,anchor=base,inner sep=0pt, outer sep=0pt, scale=  0.80] at (165.70, 41.95) {60};

\node[text=drawColor,anchor=base,inner sep=0pt, outer sep=0pt, scale=  0.80] at (207.19, 41.95) {80};
\end{scope}
\begin{scope}
\path[clip] (  0.00,  0.00) rectangle (222.67,222.67);
\definecolor{drawColor}{RGB}{0,0,0}

\node[text=drawColor,anchor=base,inner sep=0pt, outer sep=0pt, scale=  1.00] at (125.25, 31.00) {Validation Step};
\end{scope}
\begin{scope}
\path[clip] (  0.00,  0.00) rectangle (222.67,222.67);
\definecolor{drawColor}{RGB}{0,0,0}

\node[text=drawColor,rotate= 90.00,anchor=base,inner sep=0pt, outer sep=0pt, scale=  1.00] at ( 11.89,122.78) {Validation return};
\end{scope}
\end{tikzpicture}

%% file: paper.bbl
\begin{thebibliography}{10}
\expandafter\ifx\csname url\endcsname\relax
  \def\url#1{\texttt{#1}}\fi
\expandafter\ifx\csname urlprefix\endcsname\relax\def\urlprefix{URL }\fi
\expandafter\ifx\csname href\endcsname\relax
  \def\href#1#2{#2} \def\path#1{#1}\fi

\bibitem{goodfellow16}
I.~Goodfellow, Y.~Bengio, A.~Courville, Deep Learning, MIT Press, 2016,
  \url{http://www.deeplearningbook.org}.

\bibitem{murphy12}
K.~P. Murphy, Machine learning - a probabilistic perspective, Adaptive
  computation and machine learning series, {MIT} Press, 2012.

\bibitem{lecun15}
Y.~LeCun, Y.~Bengio, G.~E. Hinton,
  \href{https://doi.org/10.1038/nature14539}{Deep learning}, Nat. 521~(7553)
  (2015) 436--444.
\newblock \href {https://doi.org/10.1038/nature14539}
  {\path{doi:10.1038/nature14539}}.
\newline\urlprefix\url{https://doi.org/10.1038/nature14539}

\bibitem{Levine16}
S.~Levine, C.~Finn, T.~Darrell, P.~Abbeel,
  \href{http://jmlr.org/papers/v17/15-522.html}{End-to-end training of deep
  visuomotor policies}, J. Mach. Learn. Res. 17 (2016) 39:1--39:40.
\newline\urlprefix\url{http://jmlr.org/papers/v17/15-522.html}

\bibitem{Hoof16}
H.~van Hoof, N.~Chen, M.~Karl, P.~van~der Smagt, J.~Peters,
  \href{https://doi.org/10.1109/IROS.2016.7759578}{Stable reinforcement
  learning with autoencoders for tactile and visual data}, in: 2016 {IEEE/RSJ}
  International Conference on Intelligent Robots and Systems, {IROS} 2016,
  Daejeon, South Korea, October 9-14, 2016, {IEEE}, 2016, pp. 3928--3934.
\newblock \href {https://doi.org/10.1109/IROS.2016.7759578}
  {\path{doi:10.1109/IROS.2016.7759578}}.
\newline\urlprefix\url{https://doi.org/10.1109/IROS.2016.7759578}

\bibitem{openai18}
OpenAI, M.~Andrychowicz, B.~Baker, M.~Chociej, R.~J{\'{o}}zefowicz, B.~McGrew,
  J.~W. Pachocki, J.~Pachocki, A.~Petron, M.~Plappert, G.~Powell, A.~Ray,
  J.~Schneider, S.~Sidor, J.~Tobin, P.~Welinder, L.~Weng, W.~Zaremba,
  \href{http://arxiv.org/abs/1808.00177}{Learning dexterous in-hand
  manipulation}, CoRR abs/1808.00177 (2018).
\newblock \href {http://arxiv.org/abs/1808.00177} {\path{arXiv:1808.00177}}.
\newline\urlprefix\url{http://arxiv.org/abs/1808.00177}

\bibitem{Kalashnikov18}
D.~Kalashnikov, A.~Irpan, P.~Pastor, J.~Ibarz, A.~Herzog, E.~Jang, D.~Quillen,
  E.~Holly, M.~Kalakrishnan, V.~Vanhoucke, S.~Levine,
  \href{http://arxiv.org/abs/1806.10293}{Qt-opt: Scalable deep reinforcement
  learning for vision-based robotic manipulation}, CoRR abs/1806.10293 (2018).
\newblock \href {http://arxiv.org/abs/1806.10293} {\path{arXiv:1806.10293}}.
\newline\urlprefix\url{http://arxiv.org/abs/1806.10293}

\bibitem{Bhalla20}
S.~Bhalla, S.~G. Subramanian, M.~Crowley,
  \href{https://doi.org/10.1007/978-3-030-47358-7\_7}{Deep multi agent
  reinforcement learning for autonomous driving}, in: C.~Goutte, X.~Zhu (Eds.),
  Advances in Artificial Intelligence - 33rd Canadian Conference on Artificial
  Intelligence, Canadian {AI} 2020, Ottawa, ON, Canada, May 13-15, 2020,
  Proceedings, Vol. 12109 of Lecture Notes in Computer Science, Springer, 2020,
  pp. 67--78.
\newblock \href {https://doi.org/10.1007/978-3-030-47358-7\_7}
  {\path{doi:10.1007/978-3-030-47358-7\_7}}.
\newline\urlprefix\url{https://doi.org/10.1007/978-3-030-47358-7\_7}

\bibitem{Baheri20}
A.~Baheri, S.~Nageshrao, H.~E. Tseng, I.~V. Kolmanovsky, A.~Girard, D.~P.
  Filev, \href{https://doi.org/10.1109/IV47402.2020.9304744}{Deep reinforcement
  learning with enhanced safety for autonomous highway driving}, in: {IEEE}
  Intelligent Vehicles Symposium, {IV} 2020, Las Vegas, NV, USA, October 19 -
  November 13, 2020, {IEEE}, 2020, pp. 1550--1555.
\newblock \href {https://doi.org/10.1109/IV47402.2020.9304744}
  {\path{doi:10.1109/IV47402.2020.9304744}}.
\newline\urlprefix\url{https://doi.org/10.1109/IV47402.2020.9304744}

\bibitem{Huang21}
Z.~Huang, J.~Wu, C.~Lv, \href{https://arxiv.org/abs/2103.10690}{Efficient deep
  reinforcement learning with imitative expert priors for autonomous driving},
  CoRR abs/2103.10690 (2021).
\newblock \href {http://arxiv.org/abs/2103.10690} {\path{arXiv:2103.10690}}.
\newline\urlprefix\url{https://arxiv.org/abs/2103.10690}

\bibitem{silver16}
D.~Silver, A.~Huang, C.~J. Maddison, A.~Guez, L.~Sifre, G.~van~den Driessche,
  J.~Schrittwieser, I.~Antonoglou, V.~Panneershelvam, M.~Lanctot, S.~Dieleman,
  D.~Grewe, J.~Nham, N.~Kalchbrenner, I.~Sutskever, T.~Lillicrap, M.~Leach,
  K.~Kavukcuoglu, T.~Graepel, D.~Hassabis, Mastering the game of {Go} with deep
  neural networks and tree search, Nature 529~(7587) (2016) 484--489.
\newblock \href {https://doi.org/10.1038/nature16961}
  {\path{doi:10.1038/nature16961}}.

\bibitem{silver17a}
D.~Silver, J.~Schrittwieser, K.~Simonyan, I.~Antonoglou, A.~Huang, A.~Guez,
  T.~Hubert, L.~Baker, M.~Lai, A.~Bolton, Y.~Chen, T.~Lillicrap, F.~Hui,
  L.~Sifre, G.~van~den Driessche, T.~Graepel, D.~Hassabis,
  \href{http://dx.doi.org/10.1038/nature24270}{Mastering the game of go without
  human knowledge}, Nature 550 (2017) 354--.
\newblock \href {https://doi.org/10.1038/nature24270}
  {\path{doi:10.1038/nature24270}}.
\newline\urlprefix\url{http://dx.doi.org/10.1038/nature24270}

\bibitem{silver17b}
D.~Silver, T.~Hubert, J.~Schrittwieser, I.~Antonoglou, M.~Lai, A.~Guez,
  M.~Lanctot, L.~Sifre, D.~Kumaran, T.~Graepel, T.~P. Lillicrap, K.~Simonyan,
  D.~Hassabis, \href{http://arxiv.org/abs/1712.01815}{Mastering chess and shogi
  by self-play with a general reinforcement learning algorithm}, CoRR
  abs/1712.01815 (2017).
\newblock \href {http://arxiv.org/abs/1712.01815} {\path{arXiv:1712.01815}}.
\newline\urlprefix\url{http://arxiv.org/abs/1712.01815}

\bibitem{Schrittwieser19}
J.~Schrittwieser, I.~Antonoglou, T.~Hubert, K.~Simonyan, L.~Sifre, S.~Schmitt,
  A.~Guez, E.~Lockhart, D.~Hassabis, T.~Graepel, T.~P. Lillicrap, D.~Silver,
  \href{http://arxiv.org/abs/1911.08265}{Mastering atari, go, chess and shogi
  by planning with a learned model}, CoRR abs/1911.08265 (2019).
\newblock \href {http://arxiv.org/abs/1911.08265} {\path{arXiv:1911.08265}}.
\newline\urlprefix\url{http://arxiv.org/abs/1911.08265}

\bibitem{Badia20}
A.~P. Badia, B.~Piot, S.~Kapturowski, P.~Sprechmann, A.~Vitvitskyi, D.~Guo,
  C.~Blundell, \href{https://arxiv.org/abs/2003.13350}{Agent57: Outperforming
  the atari human benchmark}, CoRR abs/2003.13350 (2020).
\newblock \href {http://arxiv.org/abs/2003.13350} {\path{arXiv:2003.13350}}.
\newline\urlprefix\url{https://arxiv.org/abs/2003.13350}

\bibitem{ale}
M.~G. Bellemare, Y.~Naddaf, J.~Veness, M.~Bowling,
  \href{http://arxiv.org/abs/1207.4708}{The arcade learning environment: An
  evaluation platform for general agents}, CoRR abs/1207.4708 (2012).
\newblock \href {http://arxiv.org/abs/1207.4708} {\path{arXiv:1207.4708}}.
\newline\urlprefix\url{http://arxiv.org/abs/1207.4708}

\bibitem{doubledqn}
H.~van Hasselt, A.~Guez, D.~Silver, \href{http://arxiv.org/abs/1509.06461}{Deep
  reinforcement learning with double q-learning}, CoRR abs/1509.06461 (2015).
\newblock \href {http://arxiv.org/abs/1509.06461} {\path{arXiv:1509.06461}}.
\newline\urlprefix\url{http://arxiv.org/abs/1509.06461}

\bibitem{Hasselt18}
H.~van Hasselt, Y.~Doron, F.~Strub, M.~Hessel, N.~Sonnerat, J.~Modayil,
  \href{http://arxiv.org/abs/1812.02648}{Deep reinforcement learning and the
  deadly triad}, CoRR abs/1812.02648 (2018).
\newblock \href {http://arxiv.org/abs/1812.02648} {\path{arXiv:1812.02648}}.
\newline\urlprefix\url{http://arxiv.org/abs/1812.02648}

\bibitem{Ilyas18}
A.~Ilyas, L.~Engstrom, S.~Santurkar, D.~Tsipras, F.~Janoos, L.~Rudolph,
  A.~Madry, \href{http://arxiv.org/abs/1811.02553}{Are deep policy gradient
  algorithms truly policy gradient algorithms?}, CoRR abs/1811.02553 (2018).
\newblock \href {http://arxiv.org/abs/1811.02553} {\path{arXiv:1811.02553}}.
\newline\urlprefix\url{http://arxiv.org/abs/1811.02553}

\bibitem{Agarwal19}
A.~Agarwal, S.~M. Kakade, J.~D. Lee, G.~Mahajan,
  \href{http://arxiv.org/abs/1908.00261}{Optimality and approximation with
  policy gradient methods in markov decision processes}, CoRR abs/1908.00261
  (2019).
\newblock \href {http://arxiv.org/abs/1908.00261} {\path{arXiv:1908.00261}}.
\newline\urlprefix\url{http://arxiv.org/abs/1908.00261}

\bibitem{nielsen16}
M.~A. Nielsen, I.~L. Chuang,
  \href{https://www.cambridge.org/de/academic/subjects/physics/quantum-physics-quantum-information-and-quantum-computation/quantum-computation-and-quantum-information-10th-anniversary-edition?format=HB}{Quantum
  Computation and Quantum Information (10th Anniversary edition)}, Cambridge
  University Press, 2016.
\newline\urlprefix\url{https://www.cambridge.org/de/academic/subjects/physics/quantum-physics-quantum-information-and-quantum-computation/quantum-computation-and-quantum-information-10th-anniversary-edition?format=HB}

\bibitem{Shor99}
P.~W. Shor, \href{https://doi.org/10.1137/S0036144598347011}{Polynomial-time
  algorithms for prime factorization and discrete logarithms on a quantum
  computer}, {SIAM} Rev. 41~(2) (1999) 303--332.
\newblock \href {https://doi.org/10.1137/S0036144598347011}
  {\path{doi:10.1137/S0036144598347011}}.
\newline\urlprefix\url{https://doi.org/10.1137/S0036144598347011}

\bibitem{Grover96}
L.~K. Grover, \href{https://doi.org/10.1145/237814.237866}{A fast quantum
  mechanical algorithm for database search}, in: G.~L. Miller (Ed.),
  Proceedings of the Twenty-Eighth Annual {ACM} Symposium on the Theory of
  Computing, Philadelphia, Pennsylvania, USA, May 22-24, 1996, {ACM}, 1996, pp.
  212--219.
\newblock \href {https://doi.org/10.1145/237814.237866}
  {\path{doi:10.1145/237814.237866}}.
\newline\urlprefix\url{https://doi.org/10.1145/237814.237866}

\bibitem{Grover98}
L.~K. Grover, \href{https://doi.org/10.1145/276698.276712}{A framework for fast
  quantum mechanical algorithms}, in: J.~S. Vitter (Ed.), Proceedings of the
  Thirtieth Annual {ACM} Symposium on the Theory of Computing, Dallas, Texas,
  USA, May 23-26, 1998, {ACM}, 1998, pp. 53--62.
\newblock \href {https://doi.org/10.1145/276698.276712}
  {\path{doi:10.1145/276698.276712}}.
\newline\urlprefix\url{https://doi.org/10.1145/276698.276712}

\bibitem{StilckFranca2021}
D.~Stilck~Fran{\c{c}}a, R.~Garc{\'i}a-Patr{\'o}n,
  \href{https://doi.org/10.1038/s41567-021-01356-3}{Limitations of optimization
  algorithms on noisy quantum devices}, Nature Physics 17~(11) (2021)
  1221--1227.
\newblock \href {https://doi.org/10.1038/s41567-021-01356-3}
  {\path{doi:10.1038/s41567-021-01356-3}}.
\newline\urlprefix\url{https://doi.org/10.1038/s41567-021-01356-3}

\bibitem{Buhrman:2021}
H.~Buhrman, B.~Loff, S.~Patro, F.~Speelman,
  \href{https://arxiv.org/abs/2106.02005}{Limits of quantum speed-ups for
  computational geometry and other problems: Fine-grained complexity via
  quantum walks}, CoRR abs/2106.02005 (2021).
\newblock \href {http://arxiv.org/abs/2106.02005} {\path{arXiv:2106.02005}}.
\newline\urlprefix\url{https://arxiv.org/abs/2106.02005}

\bibitem{Preskill2018quantumcomputingin}
J.~Preskill, \href{https://doi.org/10.22331/q-2018-08-06-79}{Quantum
  {C}omputing in the {NISQ} era and beyond}, {Quantum} 2 (2018) 79.
\newblock \href {https://doi.org/10.22331/q-2018-08-06-79}
  {\path{doi:10.22331/q-2018-08-06-79}}.
\newline\urlprefix\url{https://doi.org/10.22331/q-2018-08-06-79}

\bibitem{bayerstadler2021}
A.~Bayerstadler, G.~Becquin, J.~Binder, T.~Botter, H.~Ehm, T.~Ehmer,
  M.~Erdmann, N.~Gaus, P.~Harbach, M.~Hess, J.~Klepsch, M.~Leib, S.~Luber,
  A.~Luckow, M.~Mansky, W.~Mauerer, F.~Neukart, C.~Niedermeier, L.~Palackal,
  R.~Pfeiffer, C.~Polenz, J.~Sepulveda, T.~Sievers, B.~Standen, M.~Streif,
  T.~Strohm, C.~Utschig-Utschig, D.~Volz, H.~Weiss, F.~Winter, Q.~Technology,
  A.~C. QUTAC, \href{https://doi.org/10.1140/epjqt/s40507-021-00114-x}{Industry
  quantum computing applications}, EPJ Quantum Technology 8~(1) (2021) 25.
\newblock \href {https://doi.org/10.1140/epjqt/s40507-021-00114-x}
  {\path{doi:10.1140/epjqt/s40507-021-00114-x}}.
\newline\urlprefix\url{https://doi.org/10.1140/epjqt/s40507-021-00114-x}

\bibitem{bova2021}
F.~Bova, A.~Goldfarb, R.~G. Melko,
  \href{https://doi.org/10.1140/epjqt/s40507-021-00091-1}{Commercial
  applications of quantum computing}, EPJ Quantum Technology 8~(1) (2021) 2.
\newblock \href {https://doi.org/10.1140/epjqt/s40507-021-00091-1}
  {\path{doi:10.1140/epjqt/s40507-021-00091-1}}.
\newline\urlprefix\url{https://doi.org/10.1140/epjqt/s40507-021-00091-1}

\bibitem{vqdqn}
S.~Y.~C. {Chen}, C.~H.~H. {Yang}, J.~{Qi}, P.~Y. {Chen}, X.~{Ma}, H.~S. {Goan},
  Variational quantum circuits for deep reinforcement learning, IEEE Access 8
  (2020) 141007--141024.
\newblock \href {https://doi.org/10.1109/ACCESS.2020.3010470}
  {\path{doi:10.1109/ACCESS.2020.3010470}}.

\bibitem{Lockwood2020}
O.~Lockwood, M.~Si, \href{https://arxiv.org/abs/2008.07524}{Reinforcement
  learning with quantum variational circuits}, CoRR abs/2008.07524 (2020).
\newblock \href {http://arxiv.org/abs/2008.07524} {\path{arXiv:2008.07524}}.
\newline\urlprefix\url{https://arxiv.org/abs/2008.07524}

\bibitem{dqn13}
V.~Mnih, K.~Kavukcuoglu, D.~Silver, A.~Graves, I.~Antonoglou, D.~Wierstra,
  M.~A. Riedmiller, \href{http://arxiv.org/abs/1312.5602}{Playing atari with
  deep reinforcement learning}, CoRR abs/1312.5602 (2013).
\newblock \href {http://arxiv.org/abs/1312.5602} {\path{arXiv:1312.5602}}.
\newline\urlprefix\url{http://arxiv.org/abs/1312.5602}

\bibitem{dqn15}
V.~Mnih, K.~Kavukcuoglu, D.~Silver, A.~A. Rusu, J.~Veness, M.~G. Bellemare,
  A.~Graves, M.~Riedmiller, A.~K. Fidjeland, G.~Ostrovski, S.~Petersen,
  C.~Beattie, A.~Sadik, I.~Antonoglou, H.~King, D.~Kumaran, D.~Wierstra,
  S.~Legg, D.~Hassabis,
  \href{http://dx.doi.org/10.1038/nature14236}{Human-level control through deep
  reinforcement learning}, Nature 518~(7540) (2015) 529--533.
\newline\urlprefix\url{http://dx.doi.org/10.1038/nature14236}

\bibitem{Mitarai18}
K.~Mitarai, M.~Negoro, M.~Kitagawa, K.~Fujii,
  \href{https://link.aps.org/doi/10.1103/PhysRevA.98.032309}{Quantum circuit
  learning}, Phys. Rev. A 98 (2018) 032309.
\newblock \href {https://doi.org/10.1103/PhysRevA.98.032309}
  {\path{doi:10.1103/PhysRevA.98.032309}}.
\newline\urlprefix\url{https://link.aps.org/doi/10.1103/PhysRevA.98.032309}

\bibitem{Broughton2021}
M.~Broughton, G.~Verdon, T.~McCourt, A.~J. Martinez, J.~H. Yoo, S.~V. Isakov,
  P.~Massey, R.~Halavati, M.~Y. Niu, A.~Zlokapa, E.~Peters, O.~Lockwood,
  A.~Skolik, S.~Jerbi, V.~Dunjko, M.~Leib, M.~Streif, D.~V. Dollen, H.~Chen,
  S.~Cao, R.~Wiersema, H.-Y. Huang, J.~R. McClean, R.~Babbush, S.~Boixo,
  D.~Bacon, A.~K. Ho, H.~Neven, M.~Mohseni, Tensorflow quantum: A software
  framework for quantum machine learning (2021).
\newblock \href {http://arxiv.org/abs/2003.02989} {\path{arXiv:2003.02989}}.

\bibitem{Abadi15}
M.~Abadi, A.~Agarwal, P.~Barham, E.~Brevdo, Z.~Chen, C.~Citro, G.~S. Corrado,
  A.~Davis, J.~Dean, M.~Devin, S.~Ghemawat, I.~Goodfellow, A.~Harp, G.~Irving,
  M.~Isard, Y.~Jia, R.~Jozefowicz, L.~Kaiser, M.~Kudlur, J.~Levenberg,
  D.~Man\'{e}, R.~Monga, S.~Moore, D.~Murray, C.~Olah, M.~Schuster, J.~Shlens,
  B.~Steiner, I.~Sutskever, K.~Talwar, P.~Tucker, V.~Vanhoucke, V.~Vasudevan,
  F.~Vi\'{e}gas, O.~Vinyals, P.~Warden, M.~Wattenberg, M.~Wicke, Y.~Yu,
  X.~Zheng, \href{https://www.tensorflow.org/}{{TensorFlow}: Large-scale
  machine learning on heterogeneous systems}, software available from
  tensorflow.org (2015).
\newline\urlprefix\url{https://www.tensorflow.org/}

\bibitem{Abraham19}
H.~\text{Abraham et al.}, Qiskit: An open-source framework for quantum
  computing (2019).
\newblock \href {https://doi.org/10.5281/zenodo.2562110}
  {\path{doi:10.5281/zenodo.2562110}}.

\bibitem{Paszke2019}
A.~Paszke, S.~Gross, F.~Massa, A.~Lerer, J.~Bradbury, G.~Chanan, T.~Killeen,
  Z.~Lin, N.~Gimelshein, L.~Antiga, A.~Desmaison, A.~Kopf, E.~Yang, Z.~DeVito,
  M.~Raison, A.~Tejani, S.~Chilamkurthy, B.~Steiner, L.~Fang, J.~Bai,
  S.~Chintala,
  \href{http://papers.neurips.cc/paper/9015-pytorch-an-imperative-style-high-performance-deep-learning-library.pdf}{Pytorch:
  An imperative style, high-performance deep learning library}, in: H.~Wallach,
  H.~Larochelle, A.~Beygelzimer, F.~d\textquotesingle Alch\'{e}-Buc, E.~Fox,
  R.~Garnett (Eds.), Advances in Neural Information Processing Systems 32,
  Curran Associates, Inc., 2019, pp. 8024--8035.
\newline\urlprefix\url{http://papers.neurips.cc/paper/9015-pytorch-an-imperative-style-high-performance-deep-learning-library.pdf}

\bibitem{Sutton99}
R.~S. Sutton, D.~McAllester, S.~Singh, Y.~Mansour, Policy gradient methods for
  reinforcement learning with function approximation, in: Proceedings of the
  12th International Conference on Neural Information Processing Systems,
  NIPS'99, MIT Press, Cambridge, MA, USA, 1999, p. 1057–1063.

\bibitem{Mnih16}
V.~Mnih, A.~P. Badia, M.~Mirza, A.~Graves, T.~P. Lillicrap, T.~Harley,
  D.~Silver, K.~Kavukcuoglu,
  \href{http://arxiv.org/abs/1602.01783}{Asynchronous methods for deep
  reinforcement learning}, CoRR abs/1602.01783 (2016).
\newblock \href {http://arxiv.org/abs/1602.01783} {\path{arXiv:1602.01783}}.
\newline\urlprefix\url{http://arxiv.org/abs/1602.01783}

\bibitem{Lin92}
L.~J. Lin, \href{https://doi.org/10.1007/BF00992699}{Self-improving reactive
  agents based on reinforcement learning, planning and teaching}, Mach. Learn.
  8 (1992) 293--321.
\newblock \href {https://doi.org/10.1007/BF00992699}
  {\path{doi:10.1007/BF00992699}}.
\newline\urlprefix\url{https://doi.org/10.1007/BF00992699}

\bibitem{arute2019quantum}
F.~Arute, K.~Arya, R.~Babbush, D.~Bacon, J.~C. Bardin, R.~Barends, R.~Biswas,
  S.~Boixo, F.~G. Brandao, D.~A. Buell, et~al., Quantum supremacy using a
  programmable superconducting processor, Nature 574~(7779) (2019) 505--510.
\newblock \href {https://doi.org/10.1038/s41586-019-1666-5}
  {\path{doi:10.1038/s41586-019-1666-5}}.

\bibitem{skolik2021}
A.~Skolik, S.~Jerbi, V.~Dunjko, Quantum agents in the gym: a variational
  quantum algorithm for deep q-learning, arXiv preprint arXiv:2103.15084
  (2021).

\bibitem{Brockman16}
G.~Brockman, V.~Cheung, L.~Pettersson, J.~Schneider, J.~Schulman, J.~Tang,
  W.~Zaremba, \href{http://arxiv.org/abs/1606.01540}{Openai gym}, CoRR
  abs/1606.01540 (2016).
\newblock \href {http://arxiv.org/abs/1606.01540} {\path{arXiv:1606.01540}}.
\newline\urlprefix\url{http://arxiv.org/abs/1606.01540}

\bibitem{WatkinsDayan92}
C.~J. C.~H. Watkins, P.~Dayan,
  \href{https://doi.org/10.1007/BF00992698}{Q-learning}, Machine Learning 8~(3)
  (1992) 279--292.
\newblock \href {https://doi.org/10.1007/BF00992698}
  {\path{doi:10.1007/BF00992698}}.
\newline\urlprefix\url{https://doi.org/10.1007/BF00992698}

\bibitem{duelingdqn}
Z.~Wang, N.~de~Freitas, M.~Lanctot,
  \href{http://arxiv.org/abs/1511.06581}{Dueling network architectures for deep
  reinforcement learning}, CoRR abs/1511.06581 (2015).
\newblock \href {http://arxiv.org/abs/1511.06581} {\path{arXiv:1511.06581}}.
\newline\urlprefix\url{http://arxiv.org/abs/1511.06581}

\bibitem{prioritizedreplay}
T.~Schaul, J.~Quan, I.~Antonoglou, D.~Silver,
  \href{http://arxiv.org/abs/1511.05952}{Prioritized experience replay}, in:
  Y.~Bengio, Y.~LeCun (Eds.), 4th International Conference on Learning
  Representations, {ICLR} 2016, San Juan, Puerto Rico, May 2-4, 2016,
  Conference Track Proceedings, 2016.
\newline\urlprefix\url{http://arxiv.org/abs/1511.05952}

\bibitem{Horgan18}
D.~Horgan, J.~Quan, D.~Budden, G.~Barth{-}Maron, M.~Hessel, H.~van Hasselt,
  D.~Silver, \href{https://openreview.net/forum?id=H1Dy---0Z}{Distributed
  prioritized experience replay}, in: 6th International Conference on Learning
  Representations, {ICLR} 2018, Vancouver, BC, Canada, April 30 - May 3, 2018,
  Conference Track Proceedings, OpenReview.net, 2018.
\newline\urlprefix\url{https://openreview.net/forum?id=H1Dy---0Z}

\bibitem{gorilladqn}
A.~Nair, P.~Srinivasan, S.~Blackwell, C.~Alcicek, R.~Fearon, A.~D. Maria,
  V.~Panneershelvam, M.~Suleyman, C.~Beattie, S.~Petersen, S.~Legg, V.~Mnih,
  K.~Kavukcuoglu, D.~Silver, \href{http://arxiv.org/abs/1507.04296}{Massively
  parallel methods for deep reinforcement learning}, CoRR abs/1507.04296
  (2015).
\newblock \href {http://arxiv.org/abs/1507.04296} {\path{arXiv:1507.04296}}.
\newline\urlprefix\url{http://arxiv.org/abs/1507.04296}

\bibitem{ngu}
A.~P. Badia, P.~Sprechmann, A.~Vitvitskyi, D.~Guo, B.~Piot, S.~Kapturowski,
  O.~Tieleman, M.~Arjovsky, A.~Pritzel, A.~Bolt, C.~Blundell,
  \href{https://openreview.net/forum?id=Sye57xStvB}{Never give up: Learning
  directed exploration strategies}, in: 8th International Conference on
  Learning Representations, {ICLR} 2020, Addis Ababa, Ethiopia, April 26-30,
  2020, OpenReview.net, 2020.
\newline\urlprefix\url{https://openreview.net/forum?id=Sye57xStvB}

\bibitem{r2d2}
S.~Kapturowski, G.~Ostrovski, J.~Quan, R.~Munos, W.~Dabney,
  \href{https://openreview.net/forum?id=r1lyTjAqYX}{Recurrent experience replay
  in distributed reinforcement learning}, in: 7th International Conference on
  Learning Representations, {ICLR} 2019, New Orleans, LA, USA, May 6-9, 2019,
  OpenReview.net, 2019.
\newline\urlprefix\url{https://openreview.net/forum?id=r1lyTjAqYX}

\bibitem{Tsitsiklis97}
J.~N. {Tsitsiklis}, B.~{Van Roy}, An analysis of temporal-difference learning
  with function approximation, IEEE Transactions on Automatic Control 42~(5)
  (1997) 674--690.
\newblock \href {https://doi.org/10.1109/9.580874}
  {\path{doi:10.1109/9.580874}}.

\bibitem{Hasselt10}
H.~van Hasselt,
  \href{https://proceedings.neurips.cc/paper/2010/hash/091d584fced301b442654dd8c23b3fc9-Abstract.html}{Double
  q-learning}, in: J.~D. Lafferty, C.~K.~I. Williams, J.~Shawe{-}Taylor, R.~S.
  Zemel, A.~Culotta (Eds.), Advances in Neural Information Processing Systems
  23: 24th Annual Conference on Neural Information Processing Systems 2010.
  Proceedings of a meeting held 6-9 December 2010, Vancouver, British Columbia,
  Canada, Curran Associates, Inc., 2010, pp. 2613--2621.
\newline\urlprefix\url{https://proceedings.neurips.cc/paper/2010/hash/091d584fced301b442654dd8c23b3fc9-Abstract.html}

\bibitem{Sutton15}
R.~S. Sutton, A.~R. Mahmood, M.~White,
  \href{http://arxiv.org/abs/1503.04269}{An emphatic approach to the problem of
  off-policy temporal-difference learning}, CoRR abs/1503.04269 (2015).
\newblock \href {http://arxiv.org/abs/1503.04269} {\path{arXiv:1503.04269}}.
\newline\urlprefix\url{http://arxiv.org/abs/1503.04269}

\bibitem{Dong08}
D.~Dong, C.~Chen, H.~Li, T.~J. Tarn,
  \href{https://doi.org/10.1109/TSMCB.2008.925743}{Quantum reinforcement
  learning}, {IEEE} Trans. Syst. Man Cybern. Part {B} 38~(5) (2008) 1207--1220.
\newblock \href {https://doi.org/10.1109/TSMCB.2008.925743}
  {\path{doi:10.1109/TSMCB.2008.925743}}.
\newline\urlprefix\url{https://doi.org/10.1109/TSMCB.2008.925743}

\bibitem{Dunjko15}
V.~Dunjko, J.~M. Taylor, H.~J. Briegel,
  \href{http://arxiv.org/abs/1507.08482}{Framework for learning agents in
  quantum environments}, CoRR abs/1507.08482 (2015).
\newblock \href {http://arxiv.org/abs/1507.08482} {\path{arXiv:1507.08482}}.
\newline\urlprefix\url{http://arxiv.org/abs/1507.08482}

\bibitem{Flamini19}
F.~Flamini, A.~Hamann, S.~Jerbi, L.~M. Trenkwalder, H.~P. Nautrup, H.~J.
  Briegel, \href{http://arxiv.org/abs/1907.07503}{Photonic architecture for
  reinforcement learning}, CoRR abs/1907.07503 (2019).
\newblock \href {http://arxiv.org/abs/1907.07503} {\path{arXiv:1907.07503}}.
\newline\urlprefix\url{http://arxiv.org/abs/1907.07503}

\bibitem{Neukart17}
F.~Neukart, D.~V. Dollen, C.~Seidel, G.~Compostella,
  \href{http://arxiv.org/abs/1708.09354}{Quantum-enhanced reinforcement
  learning for finite-episode games with discrete state spaces}, CoRR
  abs/1708.09354 (2017).
\newblock \href {http://arxiv.org/abs/1708.09354} {\path{arXiv:1708.09354}}.
\newline\urlprefix\url{http://arxiv.org/abs/1708.09354}

\bibitem{Silver14}
D.~Silver, G.~Lever, N.~Heess, T.~Degris, D.~Wierstra, M.~A. Riedmiller,
  \href{http://dblp.uni-trier.de/db/conf/icml/icml2014.html#SilverLHDWR14}{Deterministic
  policy gradient algorithms.}, in: ICML, Vol.~32 of JMLR Workshop and
  Conference Proceedings, JMLR.org, 2014, pp. 387--395.
\newline\urlprefix\url{http://dblp.uni-trier.de/db/conf/icml/icml2014.html#SilverLHDWR14}

\bibitem{salinas2020}
A.~Pérez-Salinas, A.~Cervera~Lierta, E.~Gil-Fuster, J.~Latorre, Data
  re-uploading for a universal quantum classifier, Quantum 4 (2020) 226.
\newblock \href {https://doi.org/10.22331/q-2020-02-06-226}
  {\path{doi:10.22331/q-2020-02-06-226}}.

\bibitem{Schuld2021}
M.~Schuld, R.~Sweke, J.~J. Meyer,
  \href{http://dx.doi.org/10.1103/PhysRevA.103.032430}{Effect of data encoding
  on the expressive power of variational quantum-machine-learning models},
  Physical Review A 103~(3) (Mar 2021).
\newblock \href {https://doi.org/10.1103/physreva.103.032430}
  {\path{doi:10.1103/physreva.103.032430}}.
\newline\urlprefix\url{http://dx.doi.org/10.1103/PhysRevA.103.032430}

\bibitem{ibmq2021}
Ibm quantum, https://quantum-computing.ibm.com/, 2021.

\bibitem{Rumelhart1986}
D.~E. Rumelhart, J.~L. McClelland, Learning Internal Representations by Error
  Propagation, 1987, pp. 318--362.

\bibitem{Mitarai2018}
K.~Mitarai, M.~Negoro, M.~Kitagawa, K.~Fujii,
  \href{http://dx.doi.org/10.1103/PhysRevA.98.032309}{Quantum circuit
  learning}, Physical Review A 98~(3) (Sep 2018).
\newblock \href {https://doi.org/10.1103/physreva.98.032309}
  {\path{doi:10.1103/physreva.98.032309}}.
\newline\urlprefix\url{http://dx.doi.org/10.1103/PhysRevA.98.032309}

\bibitem{schuld2019evaluating}
M.~Schuld, V.~Bergholm, C.~Gogolin, J.~Izaac, N.~Killoran, Evaluating analytic
  gradients on quantum hardware, Physical Review A 99~(3) (2019) 032331.

\bibitem{baker2016designing}
B.~Baker, O.~Gupta, N.~Naik, R.~Raskar, Designing neural network architectures
  using reinforcement learning, arXiv preprint arXiv:1611.02167 (2016).

\bibitem{mcclean_barren_2018}
J.~R. {McClean}, S.~Boixo, V.~N. Smelyanskiy, R.~Babbush, H.~Neven,
  \href{https://www.nature.com/articles/s41467-018-07090-4}{Barren plateaus in
  quantum neural network training landscapes} 9~(1) (2018) 4812.
\newblock \href {https://doi.org/10.1038/s41467-018-07090-4}
  {\path{doi:10.1038/s41467-018-07090-4}}.
\newline\urlprefix\url{https://www.nature.com/articles/s41467-018-07090-4}

\bibitem{skolik2021layerwise}
A.~Skolik, J.~R. McClean, M.~Mohseni, P.~van~der Smagt, M.~Leib, Layerwise
  learning for quantum neural networks, Quantum Machine Intelligence 3~(1)
  (2021) 1--11.
\newblock \href {https://doi.org/10.1007/s42484-020-00036-4}
  {\path{doi:10.1007/s42484-020-00036-4}}.

\bibitem{he2016}
K.~He, X.~Zhang, S.~Ren, J.~Sun, Deep residual learning for image recognition
  (2016) 770--778.

\bibitem{chen2020}
T.~Chen, S.~Kornblith, M.~Norouzi, G.~Hinton, A simple framework for
  contrastive learning of visual representations (2020).

\bibitem{brown2020}
T.~B. Brown, B.~Mann, N.~Ryder, M.~Subbiah, J.~Kaplan, P.~Dhariwal,
  A.~Neelakantan, P.~Shyam, G.~Sastry, A.~Askell, et~al., Language models are
  few-shot learners, arXiv preprint arXiv:2005.14165 (2020).

\bibitem{vaswani2017}
A.~Vaswani, N.~Shazeer, N.~Parmar, J.~Uszkoreit, L.~Jones, A.~N. Gomez,
  {\L}.~Kaiser, I.~Polosukhin, Attention is all you need (2017) 5998--6008.

\bibitem{Bengio2012}
Y.~Bengio, \href{https://doi.org/10.1007/978-3-642-35289-8_26}{Practical
  Recommendations for Gradient-Based Training of Deep Architectures}, Springer
  Berlin Heidelberg, Berlin, Heidelberg, 2012, pp. 437--478.
\newblock \href {https://doi.org/10.1007/978-3-642-35289-8_26}
  {\path{doi:10.1007/978-3-642-35289-8_26}}.
\newline\urlprefix\url{https://doi.org/10.1007/978-3-642-35289-8_26}

\bibitem{kingma2014}
D.~P. Kingma, J.~Ba, Adam: A method for stochastic optimization, arXiv preprint
  arXiv:1412.6980 (2014).

\bibitem{Periyasamy2022}
M.~Periyasamy, N.~Meyer, C.~Ufrecht, D.~D. Scherer, A.~Plinge, C.~Mutschler,
  \href{https://arxiv.org/abs/2205.03057}{Incremental data-uploading for
  full-quantum classification} (2022).
\newblock \href {https://doi.org/10.48550/ARXIV.2205.03057}
  {\path{doi:10.48550/ARXIV.2205.03057}}.
\newline\urlprefix\url{https://arxiv.org/abs/2205.03057}

\bibitem{Andrew2021}
A.~Wack, H.~Paik, A.~Javadi-Abhari, P.~Jurcevic, I.~Faro, J.~M. Gambetta, B.~R.
  Johnson, Quality, speed, and scale: three key attributes to measure the
  performance of near-term quantum computers (2021).
\newblock \href {http://arxiv.org/abs/2110.14108} {\path{arXiv:2110.14108}}.

\bibitem{Kakade2003}
S.~M. Kakade, On the sample complexity of reinforcement learning, Ph.D. thesis,
  Gatsby Computational Neuroscience Unit, University College London (2003).

\bibitem{Student}
Student, The probable error of a mean, Biometrika 6 (1908) 1--25.

\bibitem{Henderson2017}
P.~Henderson, R.~Islama, P.~Bachman, J.~Pineau, D.~Precup, D.~Meger,
  \href{https://arxiv.org/abs/1709.06560}{{Deep Reinforcement Learning that
  Matters}} abs/1709.06560 (2017).
\newblock \href {http://arxiv.org/abs/1709.06560} {\path{arXiv:1709.06560}}.
\newline\urlprefix\url{https://arxiv.org/abs/1709.06560}

\bibitem{Meyer2021}
N.~Meyer, {Variational Quantum Circuits for Policy Approximation}, Master's
  thesis, Friedrich-Alexander-Universit\"{a}t Erlangen-N\"{u}rnberg, Nuremberg,
  Germany (2021).

\end{thebibliography}
